\documentclass[pra,showpacs,twocolumn,aps,superscriptaddress,a4paper]{revtex4-1}
\usepackage{dcolumn,amssymb,amsmath,amsfonts,graphicx,latexsym,color}
\usepackage{epstopdf}
\usepackage{CJK}
\begin{document}

\begin{CJK*}{GBK}{song}

\title{Universal relations and normal-state properties of a Fermi gas with laser-dressed mixed-partial-wave interactions}

\author{Fang Qin}
\email{qinfang@ustc.edu.cn}
\affiliation{Key Laboratory of Quantum Information, University of Science and Technology of China, Chinese Academy of Sciences, Hefei, Anhui 230026, China}
\affiliation{Synergetic Innovation Center of Quantum Information and Quantum Physics, University of Science and Technology of China, Hefei, Anhui 230026, China}

\date{\today}

\begin{abstract}
In a recent experiment [P. Peng, $et$ $al.$, Phys. Rev. A \textbf{97}, 012702 (2018)], it has been shown that the $p$-wave Feshbach resonance can be shifted toward the $s$-wave Feshbach resonance by a laser field. Based on this experiment, we study the universal relations and the normal-state properties in an ultracold Fermi gas with coexisting $s$- and $p$-wave interactions under optical control of a $p$-wave magnetic Feshbach resonance. Within the operator-product expansion, we derive the high-momentum tail of various observable quantities in terms of contacts. We find that the high-momentum tail becomes anisotropic. Adopting the quantum virial expansion, we calculate the normal-state contacts with and without a laser field for $^{40}$K atoms using typical experimental parameters. We show that the contacts are dependent on the laser dressing. We also reveal the interplay of laser dressing and different partial-wave interactions on various contacts. In particular, we demonstrate that the impact of the laser dressing in the $p$-wave channel can be probed by measuring the $s$-wave contacts, which is a direct manifestation of few-body effects on the many-body level. Our results can be readily checked experimentally.
\end{abstract}

\maketitle

\section{Introduction}\label{1}

The interplay of $s$- and $p$-wave interactions can introduce interesting many-body physics in ultracold Fermi gases~\cite{Zhangexp2017,Zhoulihong2017,Yi2016,Yang2016,Jiang2016,Hu2016}.
Such a scenario exists in the two-component $^{40}$K Fermi gases, where the $p$-wave Feshbach resonances near $198$G are close to the wide $s$-wave Feshbach resonance near $202$G~\cite{exp2002s,exp2004s,exp2003p,review2010,exp2004p}.
In the previous studies, it has been shown that mixed-partial-wave interactions in such a system can give rise to fermion superfluid with hybridized $s$- and $p$-wave pairing~\cite{Zhoulihong2017}, as well as the interesting normal-state properties exhibiting the interplay of $s$- and $p$-wave interactions~\cite{Yi2016}.
For a low-dimensional two-component $^{40}$K Fermi gas, the overlap of $s$- and $p$-wave interactions can be tuned by using confinement-induced resonance, which would favor the elusive itinerant ferromagnetism in certain parameter regimes~\cite{Yang2016,Jiang2016,Hu2016}.

\begin{figure}
\includegraphics[width=7cm]{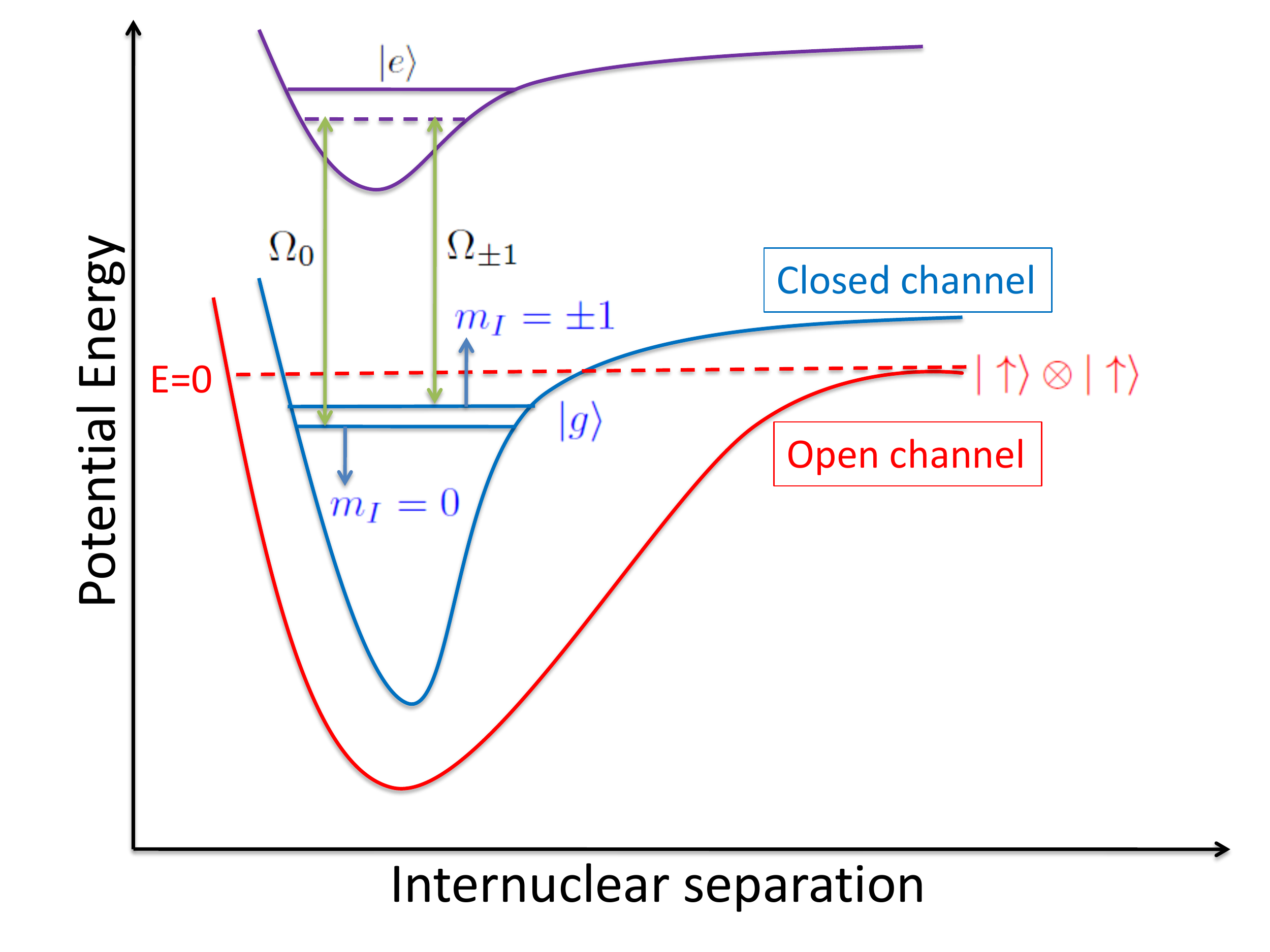}
\caption{(color online) Level scheme for the optical control of $p$-wave magnetic Feshbach resonance modulated by a laser beam~\cite{Zhangexp2017}. Here $m_I=0,\pm1$ denotes the magnetic quantum number. $\Omega_{0}$ and $\Omega_{\pm 1}$ are the effective Rabi frequencies of the laser field coupling states with $m_I=0,\pm 1$, respectively, to the excited state $|e\rangle$. } \label{fig:model}
\end{figure}

In a recent experiment~\cite{Zhangexp2017}, the $p$-wave Feshbach resonance with the magnetic quantum number $m_I=0$ is shifted
to overlap with the $s$-wave resonance via laser dressing (see Fig.~\ref{fig:model}).
As illustrated in Fig.~\ref{fig:model}, a laser field is applied to couple the bound-to-bound transition between the $p$-wave closed-channel molecular states with $m_I=0,\pm1$ and an excited state $|e\rangle$. While the energy shift is different for molecular states with different magnetic quantum numbers $m_I$, the $p$-wave Feshbach resonances associated with these closed-channel molecular states are also shifted.
The experiment thus offers an additional control on mixed-partial-wave interactions of the system, which is bound to give rise to the interesting many-body physics. As a first attempt at clarifying the impact of few-body physics on many-body properties of the system, we study the universal relations and normal-state properties in an ultracold Fermi gas with coexisting $s$- and $p$-wave interactions near a laser-dressed $p$-wave Feshbach resonance.
We expect the interplay between laser dressing and mixed-partial-wave interactions to have interesting effects on the many-body level, such that contact in the $s$-wave sector should be affected by the laser field as well. This is especially interesting as the $s$- and $p$-wave scattering channels are decoupled on the two-body level.

In dilute atomic gases with short-range interaction potentials, it has been shown that universal behaviors emerge in the large-momentum limit.
Physically, this is because when two atoms get close, the short-distance many-body wave function reduces to the two-body solution, yielding universal relations, as first studied by Tan for a three-dimensional two-component Fermi gas near an $s$-wave Feshbach resonance~\cite{Tan20081,Tan20082,Tan20083}. As a result of the universality, observable thermodynamic quantities such as the high-momentum tail of the momentum distribution, the radio-frequency (rf) spectrum, the pressure, and the energy are connected by a set of key parameters called contacts~\cite{Tan20081,Tan20082,Tan20083,Zhang2009,exp_contact1,exp_contact2,exp_contact3}.
Recently, much effort has been devoted to the study of universal relations under a synthetic gauge field~\cite{Peng2017,Jie2018,Zhang2018}, with Raman-dressed Feshbach resonance~\cite{Yi2018}, in low-dimensional atomic gases~\cite{Zwerger2011,Klumper2017,Castin20121,Castin20122,Valiente2011,Valiente2012,Zhou2017,Peng20162,Zhang2017,Yin2018,Cui20161}, in high-partial-wave quantum gases~\cite{Cui20162,Yu2015,Yu2015exp,Zhou20161,Yoshida2015,Ueda2016,Peng20161}, and in terms of contact matrices~\cite{Zhou20162} and tensors~\cite{Yoshida2016}.
In particular, the universal relations for the $p$-wave Fermi gases have already been experimentally verified~\cite{Yu2015exp}.

In this work, adopting the operator-product expansion (OPE) approach~\cite{Cui20161,Cui20162,Zhang2018,Yi2018,Wilson,Kadanoff,Braaten20081,Braaten20082,Braaten20083,Platter2016,Yu20171,Yu20172,Yu20173,Qi2016}, we derive universal relations of the system with laser-dressed hybrid interactions. We show that the leading-order terms in high-momentum tails of the momentum distribution can be expressed by five contacts, with one laser-field-dependent open-channel contact and four laser-field-dependent closed-channel contacts.
Interestingly, one of the contacts is anisotropic, and the high-momentum tail in the momentum distribution shows anisotropic features.
Notice that the anisotropy here is not due to the laser dressing, which does not induce a momentum transfer.
Rather, it comes from the anisotropy of the many-body system, due to either an anisotropic environment or spontaneous symmetry breaking~\cite{Cui20162}.

The comparisons between our results and those in previous studies are as follows:

First, for one-dimensional pure $p$-wave Fermi gases in a two-channel model of the previous work~\cite{Cui20162}, there are four contacts which are similar to our results. We will point out this in the end of the first paragraph below Eq.~(\ref{cQ2mI}). In Fermi gases with $s$-wave interactions, the leading-order term in the high-momentum tail should also feature a contact~\cite{Zwerger2011}. Therefore, when one considers a system with mixed-$s$- and $p$-wave interactions, there should be five contacts in the lead-order terms. In a previous study~\cite{Yi2016}, we also considered contacts in a system with mixed-$s$- and $p$-wave interactions, but in the absence of laser dressing. The discrepancy lies in the fact that we have previously neglected the anisotropy associated with a finite center-of-mass momentum.

Second, anisotropy in contacts can be induced either by the anisotropic interactions, such as the $p$-wave interaction or by finite center-of-mass momentum. The former has been discussed in previous works~\cite{Yu2015,Yu2015exp,Zhou20161,Yoshida2015,Ueda2016,Peng20161}, under which contacts associated with different magnetic quantum numbers $m_I$ behave differently. The latter has been discussed, for example, in Ref.~\cite{Cui20162}, where the contact in the $q^{-3}$ tail ($q$ is the relative momentum) is anisotropic, and the other contacts are isotropic. Specifically, what we mainly focus on here is the anisotropic feature induced by the center-of-mass momentum. For three-dimensional pure $p$-wave Fermi gases in Refs.~\cite{Yu2015,Yu2015exp}, they assumed that the distribution of center-of-mass momentum is isotropic, so that the contacts which they obtained do not show an anisotropic feature induced by the center-of-mass momentum and they did not have the $q^{-3}$ tail. Reference~\cite{Zhou20161} just considered the zero center-of-mass-momentum case, which does not show an anisotropic feature induced by the center-of-mass momentum. Reference~\cite{Yoshida2015} calculated only the leading order term ($q^{-2}$ tail) of the high-momentum distribution in three-dimensional pure $p$-wave Fermi gases, and it does not show an anisotropic feature induced by the center-of-mass momentum.

We then calculate the normal-state contacts and spectral function using the quantum virial expansion, both with and without the laser field for $^{40}$K atoms using typical experimental parameters. We find that, with the addition of the laser dressing in the closed channel of the $p$-wave interaction, the $s$-wave contact significantly decreases around the $p$-wave Feshbach resonance. Such a behavior is a direct manifestation of few-body effects on the many-body level and is useful for detecting the impact of dressing lasers on the system.
Furthermore, the interplay of laser dressing and $p$-wave interaction leads to a much larger $p$-wave contact than the one without a laser.
Additionally, we show the $p$-wave contacts decrease very rapidly in the Bose-Einstein condensation (BEC) limit under the influence of $s$-wave interaction, which is due to the interplay of $s$- and $p$-wave interactions on the many-body level as discussed in Ref.~\cite{Yi2016}. Our results can be readily checked under current experimental conditions.

The paper is organized as follows: In Sec.~\ref{2}, we give the model Lagrangian density to describe the two-component ultracold Fermi gas with laser coupling.
In Sec.~\ref{3}, we present a brief derivation of the renormalization of bare interactions.
In Sec.~\ref{4}, we calculate the high-momentum distribution of this system within the OPE quantum field method. In Sec.~\ref{5}, we derive the corresponding universal relations such as high-frequency rf spectroscopy, adiabatic relations, pressure relations, and virial theorem for this system. In Sec.~\ref{6}, we present the formalism of the quantum virial expansion and express the contacts and the spectral function in the normal state up to the second order. In Sec.~\ref{7}, we numerically evaluate the high-temperature contacts and spectral functions. We summarize in Sec.~\ref{8}.

\section{Model}\label{2}

We consider a two-component Femi gas close to an $s$-wave Feshbach resonance. One of the spin components is also close to a
laser-dressed $p$-wave Feshbach resonance, as illustrated in Fig.~\ref{fig:model}.
Physically, the closed-channel molecular states with different $m_I$ should feel a different laser-induced energy shift, which would lead to a state-dependent shift in the corresponding self-energies of the system Lagrangian. The local Lagrangian density (at coordinate ${\bf R}$) is given by ${\cal L} = {\cal L}_{{\rm A}} + {\cal L}_{{\rm M}} + {\cal L}_{{\rm AM}}$,
where~\cite{Zhangexp2017}
\begin{eqnarray}
{\cal L}_{{\rm A}} &=& \sum_{\sigma=\uparrow,\downarrow}\psi_{\sigma}^{\dagger}\left(i\partial_{t}+\frac{\nabla^{2}_{\bf R}}{2m}\right)\psi_{\sigma} - u_{s}\psi_{\uparrow}^{\dagger}\psi_{\downarrow}^{\dagger}\psi_{\downarrow}\psi_{\uparrow}, \label{eq:LA}\\
{\cal L}_{{\rm M}} &=& \sum_{m_I}\varphi_{m_I}^{\dagger}\left[i\partial_{t}+\frac{\nabla^{2}_{\bf R}}{4m}-\nu_{m_I}-\Sigma_{m_I}({\bf R})\right]\varphi_{m_I}, \label{eq:LM}\nonumber \\ \\
{\cal L}_{{\rm AM}} &=& -\sum_{m_I} \frac{g_{m_I}}{\sqrt{2}} \left(\varphi_{m_I}^{\dagger}{\cal Y}_{m_I} + {\cal Y}^{\dagger}_{m_I}\varphi_{m_I} \right). \label{eq:LAM}
\end{eqnarray}
Here the self-energy in coordinate space is~\cite{HePRL2018}
\begin{eqnarray}
\Sigma_{m_I}({\bf R}) = \frac{|\Omega_{m_I}|^{2}}{4\left( i\partial_{t}+\frac{\nabla^{2}_{\bf R}}{4m} - \nu_{e} + \delta_{m_I} + i\frac{\gamma_{e}}{2}\right)}, \label{eq: StarkShift}
\end{eqnarray}
\begin{eqnarray}
{\cal Y}_{m_I} = -\frac{1}{2} \sum_{\alpha} \sqrt{\frac{3}{4\pi}} C_{\alpha,m_I} \left[(i\nabla_{\alpha}\psi_{\uparrow})\psi_{\uparrow} - \psi_{\uparrow}(i\nabla_{\alpha}\psi_{\uparrow})\right], \nonumber \\
\end{eqnarray}
$\psi_{\sigma}$ ($\sigma=\uparrow,\downarrow$) denotes the
open-channel fermionic atom-field operator, $\varphi_{m_I}$ denotes the field operator for the closed-channel molecule in ground
state $|g\rangle$ with the magnetic quantum number $m_I=0,\pm1$, and
$\alpha=x,y,z$ denotes the direction of spin polarization.
$C_{\alpha,m_I}$ are the coefficients when transforming $k_{\alpha}/k$ to the $p$-wave spherical harmonics $Y_{1,m_I}({\bf \hat{k}})$, which satisfies
$\sum_{\alpha} \sqrt{3/(4\pi)} C_{\alpha,m_I} k_{\alpha} = k Y_{1,m_I}({\bf \hat{k}})$. Therefore,
$C_{x,0}=C_{y,0}=0$, $C_{z,0}=1$; $C_{x,\pm1}=\mp1/\sqrt{2}$, $C_{y,\pm1}=-i/\sqrt{2}$, $C_{z,\pm1}=0$.
${\bf R}$ is the center-of-mass coordinate, $t$ is the time, $m$ is the atom mass,
$u_{s}$ is the $s$-wave bare coupling between two fermionic atoms,
$g_{m_I}$ is the $p$-wave bare coupling between two fermionic atoms and a bosonic molecule,
and $\nu_{m_I}$ is the bare magnetic detuning.
The difference in the energy levels of atoms and excited molecules is denoted by $\nu_{e}$.
$\Omega_{m_I}$ is the strength of the effective laser-induced coupling between the molecular ground state $|g\rangle$
and excited state $|e\rangle$.
$\delta_{m_I} \equiv 2\pi (\omega_{L} - \omega_{e,m_I})$ is the detuning of the laser light with respect to the energy difference between the ground and excited states of molecules.
$\omega_{L}$ is the frequency of the laser light, and $\omega_{e,m_I}$ is the energy difference between the ground and excited states of molecules.
The spontaneous decay of the excited molecular state $|e\rangle$
is treated phenomenologically by a decay rate $\gamma_{e}$.
The natural units $\hbar=k_B=1$ will be used throughout the paper.

Accordingly, we can write the Hamiltonian in momentum space from the Lagrangian by the Legendre and Fourier transformations
\begin{align}
&H - \sum_{\sigma=\uparrow,\downarrow} \mu_{\sigma}N_{\sigma}= H_{{\rm A}} + H_{{\rm M}} + H_{{\rm AM}}, \label{eq:H}\\
&H_{{\rm A}} = \sum_{{\bf k},\sigma=\uparrow,\downarrow}a_{{\bf k},\sigma}^{\dagger} \left(\frac{k^{2}}{2m} - \mu_{\sigma} \right) a_{{\bf k},\sigma} \nonumber \\
&~~ + \frac{u_{s}}{V} \sum_{{\bf Q}',{\bf k},{\bf k}'} a_{\frac{{\bf Q}'}{2}+{\bf k},\uparrow}^{\dagger}a_{\frac{{\bf Q}'}{2}-{\bf k},\downarrow}^{\dagger}a_{\frac{{\bf Q}'}{2}-{\bf k}',\downarrow}a_{\frac{{\bf Q}'}{2}+{\bf k}',\uparrow}, \label{eq:HA} \\
&H_{{\rm M}} = \sum_{{\bf Q},m_I} b^{\dagger}_{{\bf Q},m_I}\left[\frac{Q^{2}}{4m} + \nu_{m_I} - 2\mu_{\uparrow} \right. \nonumber\\
&\left.~~ + \Sigma_{m_I}(q_{0},{\bf Q})\right]b_{{\bf Q},m_I}, \label{eq:HM}\\
&H_{{\rm AM}} = \sum_{{\bf Q},{\bf k},m_I}  \frac{g_{m_I}}{\sqrt{2V}} k \left[ Y_{1,m_I}({\bf \hat{k}}) b_{{\bf Q},m_I}^{\dagger}a_{\frac{{\bf Q}}{2}+{\bf k},\uparrow} a_{\frac{{\bf Q}}{2}-{\bf k},\uparrow} \right. \nonumber\\
&\left.~~ + Y^{*}_{1,m_I}({\bf \hat{k}}) a_{\frac{{\bf Q}}{2}-{\bf k},\uparrow}^{\dagger}a_{\frac{{\bf Q}}{2}+{\bf k},\uparrow}^{\dagger}b_{{\bf Q},m_I}\right], \label{eq:HAM}
\end{align}
where the self-energy in momentum space is
\begin{eqnarray}
\Sigma_{m_I}(q_{0},{\bf Q}) = \frac{|\Omega_{m_I}|^{2}}{4\left(q_{0} - \frac{Q^{2}}{4m} - \nu_{e} + \delta_{m_I} + i\frac{\gamma_{e}}{2}\right)}, \label{eq: StarkShift}
\end{eqnarray}
$a_{{\bf k},\sigma}$ ($a_{{\bf k},\sigma}^{\dagger}$) is the annihilation (creation) field operator of Fermi atom in momentum space, $b_{{\bf Q},m_I}$ ($b_{{\bf Q},m_I}^{\dagger}$) is the annihilation (creation) field operator of ground-state bosonic molecule in momentum space, ${\bf Q}$ is the center-of-mass momentum, $q_{0}=Q^{2}/(4m) + k^{2}/m$ is the total incoming energy, $V$ is the volume of the system, $\mu_{\sigma}$ is the fermionic chemical potential with spin $\sigma$, and the particle numbers are given by $N_{\uparrow} = \sum_{\bf k}a_{{\bf k},\uparrow}^{\dagger} a_{{\bf k},\uparrow} + 2 \sum_{{\bf Q},m_I} b_{{\bf Q},m_I}^{\dagger} b_{{\bf Q},m_I}$ and $N_{\downarrow} = \sum_{\bf k}a_{{\bf k},\downarrow}^{\dagger} a_{{\bf k},\downarrow}$.

\section{Interaction renormalization}\label{3}

In this section, we will renormalize the bare interactions in the $s$- and $p$-wave channels, respectively. On the two-body level, different partial-wave scattering channels are decoupled. Therefore, the renormalization can be performed independent for these two cases.

\subsection{$s$ wave}\label{3.1}

\begin{figure}
\includegraphics[width=7cm]{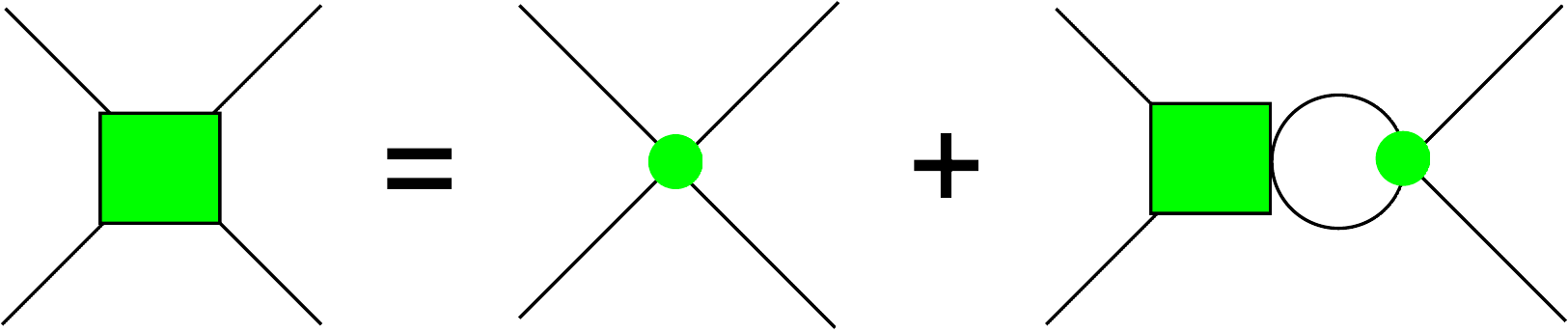}
\caption{(Color online) Diagram for calculating the $T$ matrix for $s$-wave interaction. Single lines denote the bare atom propagators $G^{(0)}$. The green square represents the $T$ matrix: $-iT_{\mathbf{k},\mathbf{k}'}^{(s)}$.
The green dot represents the interaction vertex: $-iu_{s}$.} \label{fig:Ts}
\end{figure}

In $s$-wave case, we consider
zero total momentum for each pairing state, so that an incoming state can be set as $|I_s\rangle=|{\bf k},\uparrow; - {\bf k},\downarrow\rangle$ with two fermions of different species having momentum ${\bf k}$ and $- {\bf k}$ to an outgoing state $|O_s\rangle=|{\bf k}',\uparrow; - {\bf k}',\downarrow\rangle$ with two fermions having momentum ${\bf k}'$ and $- {\bf k}'$.

As shown in Fig.~\ref{fig:Ts}, the two-body $T$ matrix for the $s$-wave interaction is given by~\cite{renormalization2007}
\begin{eqnarray}\label{eq:Ts}
-iT_{\mathbf{k},\mathbf{k}'}^{(s)}(k) = \frac{-iu_{s}}{ 1 - (-iu_{s})\Pi_{s}(k)},
\end{eqnarray}
where the polarization bubble for $s$ wave is
\begin{eqnarray}
\Pi_{s}(k)
&=& \int \frac{d^{3}{\bf p}}{(2\pi)^3} \frac{i}{k^{2}/m - p^2/m + i0^+} \nonumber\\
&=& \frac{im}{2\pi} \left( -\frac{ik}{2} - \frac{\Lambda}{\pi}\right).
\end{eqnarray}

The $s$-wave scattering length is given by
\begin{align}\label{eq:as1}
a_{s} = \frac{m}{4\pi}T_{\mathbf{k},\mathbf{k}'}^{(s)}(k=0)
= \frac{m}{4\pi}\frac{1}{\frac{1}{u_{s}} + \frac{m\Lambda}{2\pi^2}},
\end{align}
where $\Lambda$ is an ultraviolet momentum cutoff.

Further, we get the renormalization relation
\begin{eqnarray}\label{eq:as2}
\frac{1}{u_{s}}  = \frac{m}{4\pi a_{s}} - \frac{m\Lambda}{2\pi^2}.
\end{eqnarray}

\subsection{$p$ wave}\label{3.2}

\begin{figure}
\includegraphics[width=7cm]{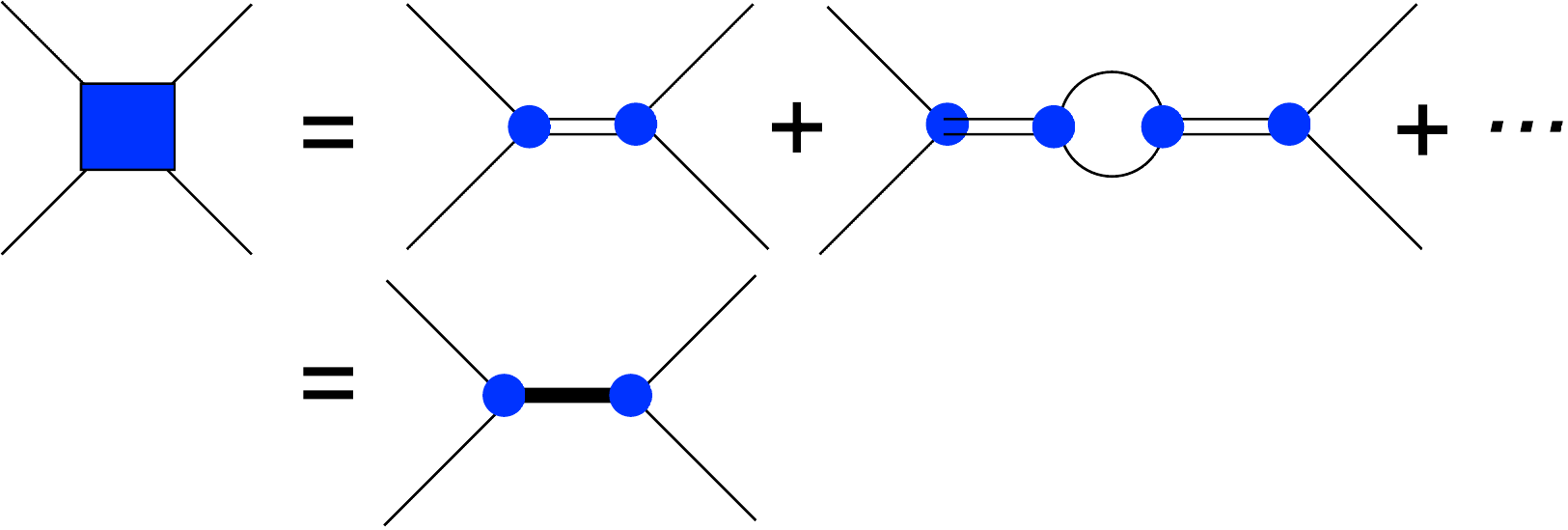}
\caption{(Color online) Diagram for calculating the $T$ matrix for $p$-wave interaction. Single lines denote the bare atom propagators $G^{(0)}$, double lines denote the bare molecule propagators $D_{m_I}^{(0)}$, and the bold one denotes the renormalized molecule propagators $D_{m_I}$. The blue square represents the $T$ matrix: $-iT_{\mathbf{k},\mathbf{k}'}^{(m_I)}$.
The blue dot represents the interaction vertex: $(-ig_{m_I}/\sqrt{2})kY_{1,m_I}({\bf \hat{k}})$.} \label{fig:Tp}
\end{figure}

We consider an incoming state $|I_p\rangle=|{\bf Q}/2 + {\bf k},\uparrow; {\bf Q}/2 - {\bf k},\uparrow\rangle$ with two fermions of different species having momentum ${\bf Q}/2 + {\bf k}$ and ${\bf Q}/2 - {\bf k}$  to an outgoing state $|O_p\rangle=|{\bf Q}/2 + {\bf k}',\uparrow; {\bf Q}/2 - {\bf k}',\uparrow\rangle$ with two fermions having momentum ${\bf Q}/2 +\mathbf{k}'$ and ${\bf Q}/2 - {\bf k}'$.

As shown in Fig.~\ref{fig:Tp}, the two-body $T$ matrix for $p$-wave interaction is given by~\cite{renormalization2007,renormalization2012,amplitude2010pwave,Yao20181,Yao20182}
\begin{widetext}\begin{eqnarray}\label{eq:Tp}
-iT_{\mathbf{k},\mathbf{k}'}^{(m_I)}(k)
&&= 2D_{m_I}^{(0)}(k)\left(\frac{-ig_{m_I}}{\sqrt{2}}\right)^2k^{2}Y_{1,m_I}({\bf \hat{k}})Y^{*}_{1,m_I}({\bf \hat{k}}') + 2D_{m_I}^{(0)2}(k)\left(\frac{-ig_{m_I}}{\sqrt{2}}\right)^{4}2\Pi_{m_I}(k)k^{2}Y_{1,m_I}({\bf \hat{k}})Y^{*}_{1,m_I}({\bf \hat{k}}') + \cdot\cdot\cdot \nonumber\\
&&= 2D_{m_I}(k)\left(\frac{-ig_{m_I}}{\sqrt{2}}\right)^2k^{2}Y_{1,m_I}({\bf \hat{k}})Y^{*}_{1,m_I}({\bf \hat{k}}'),
\end{eqnarray}\end{widetext}
where the factor 2 in front of $D_{m_I}^{(0)}(k)$ comes from the scattering of two identical fermions~\cite{Yu2015exp,renormalization2007}, the bare molecule propagator is
\begin{eqnarray}
D_{m_I}^{(0)}(k) = \frac{i}{ k^{2}/m - \nu_{m_I} - \Sigma_{m_I}(k) + i0^+},
\end{eqnarray} the polarization bubble is
\begin{eqnarray}
\Pi_{m_I}(k)
&=& \int \frac{d^{3}{\bf p}}{(2\pi)^3} \frac{ip^{2}|Y_{1,m_I}({\bf \hat{p}})|^2}{k^{2}/m - p^2/m + i0^+} \nonumber\\
&=& \frac{i}{4\pi}\left( - \frac{m\Lambda^3}{6\pi^2} - \frac{m\Lambda k^2}{2\pi^2} - \frac{im k^3}{4\pi} \right),
\end{eqnarray}
and the full molecule propagator $D_{m_I}(k)$ satisfies
\begin{eqnarray}
D_{m_I}^{-1}(k) = [D_{m_I}^{(0)}(k)]^{-1} - 2\left(\frac{-ig_{m_I}}{\sqrt{2}}\right)^2\Pi_{m_I}(k).
\end{eqnarray}

In the absence of optical field, i.e., $\Omega_{m_I}=0$, the $p$-wave scattering amplitude is given by
\begin{eqnarray}\label{eq:fp}
f_{p}(\mathbf{k},\mathbf{k}') &=& -\frac{m}{4\pi} \sum_{m_I} T^{(m_I)}_{\mathbf{k},\mathbf{k}'}(k) \nonumber\\
&=& \sum_{m_I} \frac{4\pi k^{2}Y_{1,m_I}({\bf \hat{k}})Y^{*}_{1,m_I}({\bf \hat{k}}')}{ -1/\tilde{\upsilon}_{m_I} -k^2/\tilde{R}_{m_I} - ik^3 },
\end{eqnarray} where $\tilde{\upsilon}_{m_I}$ is the $p$-wave scattering volume and $\tilde{R}_{m_I}$ is the $p$-wave effective range.
Further, we have the renormalization relations~\cite{review2010,exp2004p,Yu2015exp}
\begin{eqnarray}
\frac{\nu_{m_I}}{g_{m_I}^2} &=& \frac{\tilde{\nu}_{m_I}}{\tilde{g}_{m_I}^2} + \frac{m\Lambda^3}{24\pi^3}, \label{eq:vpRenormalization} \\
\frac{1}{g_{m_I}^2} &=& \frac{1}{\tilde{g}_{m_I}^2} - \frac{m^2\Lambda}{8\pi^3},\label{eq:RpRenormalization}
\end{eqnarray}
where $\tilde{\nu}_{m_I}/\tilde{g}_{m_I}^2$ and $1/\tilde{g}_{m_I}^2$ are renormalized in the form of
\begin{align}
&\frac{\tilde{\nu}_{m_I}}{\tilde{g}_{m_I}^2} = - \frac{m }{16\pi^2\tilde{\upsilon}_{m_I} }, \label{eq:gvp0}\\
&\frac{1}{\tilde{g}_{m_I}^2} = \frac{m^2}{ 16\pi^2\tilde{R}_{m_I}}. \label{eq:gRp0}
\end{align}

In the presence of optical field, the $p$-wave scattering volume is
\begin{eqnarray}\label{eq:vp}
\frac{1}{\upsilon_{m_I}} =  - \frac{16\pi^2 }{mg_{m_I}^2} \left[\nu_{m_I} - \frac{|\Omega_{m_I}|^{2}}{4\left( \nu_{e} - \delta_{m_I} - i\frac{\gamma_{e}}{2}\right)} \right] + \frac{2\Lambda^3}{3\pi},  \nonumber\\
\end{eqnarray}
and the $p$-wave effective range is
\begin{eqnarray}\label{eq:Rp}
\frac{1}{R_{m_I}} = \frac{16\pi^2}{m^2g_{m_I}^2} \left[ 1 + \frac{|\Omega_{m_I}|^{2}}{4\left( \nu_{e} - \delta_{m_I} - i\frac{\gamma_{e}}{2}\right)^2} \right] + \frac{2\Lambda}{\pi}. \nonumber\\
\end{eqnarray}
Notice that, in the section of numerical calculations, we use a large detuning, i.e., $\nu_{e} \ll \delta_{m_I}$~\cite{Zhangexp2017}.

\section{Momentum distribution}\label{4}

In this section, we study the tail of the momentum distribution for fermions with coexisting $s$- and $p$-wave interactions near a laser-dressed $p$-wave Feshbach resonance using the quantum field method of OPE~\cite{Cui20161,Cui20162,Zhang2018,Yi2018,Wilson,Kadanoff,Braaten20081,Braaten20082,Braaten20083,Platter2016,Yu20171,Yu20172,Yu20173,Qi2016}.

OPE is an ideal tool to explore short-range physics.
Furthermore, OPE is an operator relation that the product of two operators at small separation can be expanded in terms of the separation distance and operators, which can be interpreted as a Taylor expansion for the matrix elements of an operator. Therefore, one can expand the product of two operators as
\begin{equation}\label{ope}
\psi_\sigma^{\dag}({\bf R}-\frac{{\bf r}}{2}) \psi_\sigma({\bf R}+\frac{{\bf r}}{2}) = \sum_n C_n({\bf r}) {\cal O}_n({\bf R}),
\end{equation}
where ${\cal O}_n({\bf R})$ are the local operators and $C_n({\bf r})$ are the short-distance coefficients.
$C_n({\bf r})$ can be determined by calculating the matrix elements of the operators on both sides of Eq.~(\ref{ope}) in the two-body state $|{\bf k},\uparrow;-{\bf k},\downarrow\rangle$ for $s$-wave interaction and $|{\bf Q}/2+{\bf k},\uparrow;{\bf Q}/2-{\bf k},\uparrow\rangle$ for $p$-wave interaction.

By using the Fourier transformation on both sides of Eq.~(\ref{ope}), we have the expression of momentum distribution~\cite{Braaten20082}
\begin{align}\label{eq:rho}
& n_{\sigma}({\bf q}) =  \nonumber\\
& \int \frac{d^3{\bf R}}{V} \int d^3{\bf r} e^{-i{\bf q}\cdot{\bf r}} \left\langle \psi_{\sigma}^{\dag}({\bf R}-\frac{{\bf r}}{2}) \psi_{\sigma}({\bf R}+\frac{{\bf r}}{2}) \right\rangle,
\end{align} where $q$ is the relative momentum.

In the following subsections, we will show the calculations for the momentum distribution $n_{\uparrow}({\bf q})$, for instance.

\subsection{$s$-wave channel}\label{4.1}

\begin{figure}
\includegraphics[width=5cm]{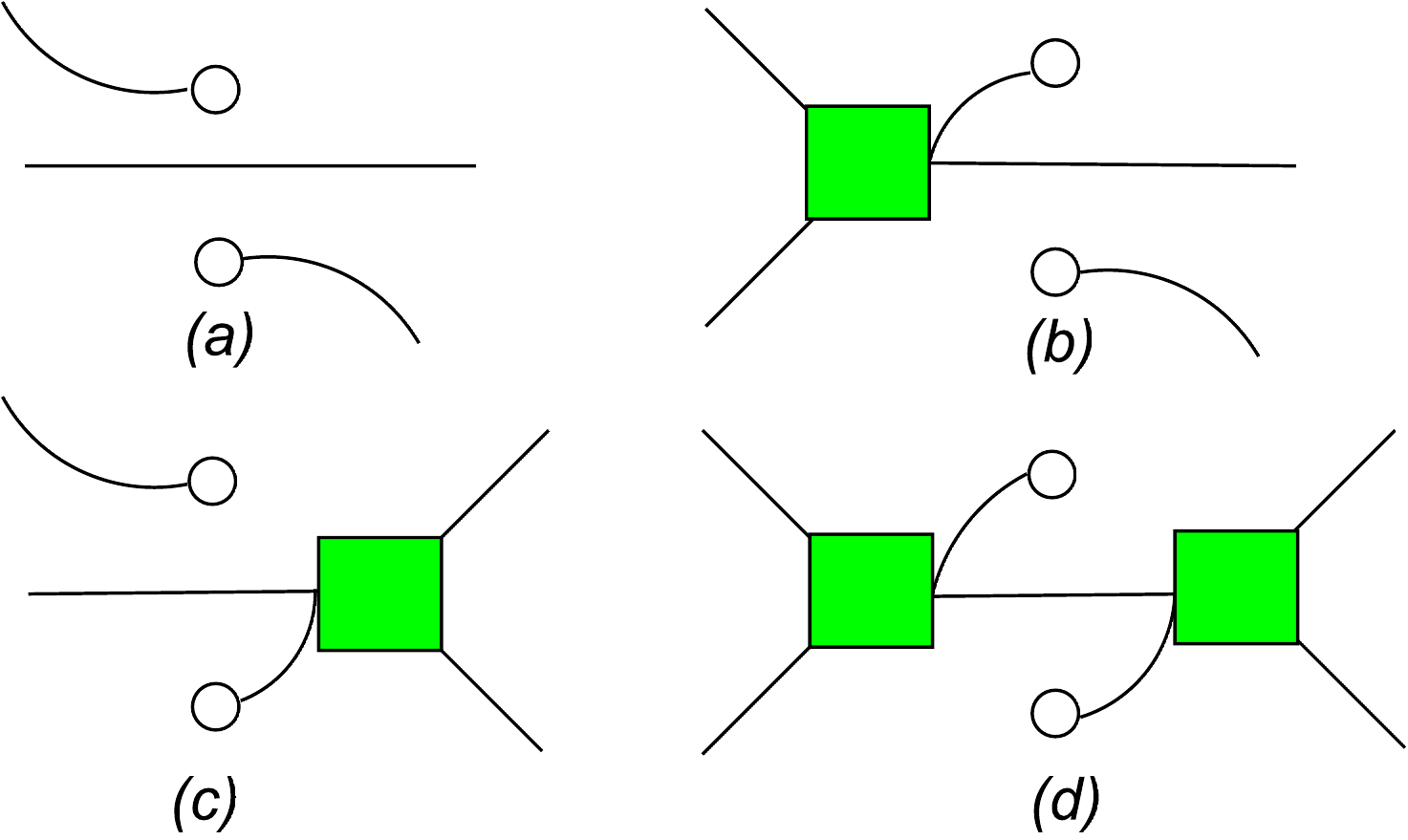}
\caption{(Color online) Diagrams for matrix elements of the
operator $\psi_\uparrow^{\dag}({\bf R}-\frac{{\bf r}}{2}) \psi_\uparrow({\bf R}+\frac{{\bf r}}{2})$ in $s$-wave interacting channel.
The open dots represent the operators. \label{fig:OneBodyOperators}}
\end{figure}

As shown in Figs.~\ref{fig:OneBodyOperators}(a)-\ref{fig:OneBodyOperators}(d), there are four types of diagrams which can be used to denote the operators on the left-hand side of OPE equation~(\ref{ope}).
However, the only nonanalyticity comes from the diagram as shown in Fig.~\ref{fig:OneBodyOperators}(d).
Therefore, we can evaluate the diagram in Fig.~\ref{fig:OneBodyOperators}(d) as
\begin{widetext}
\begin{eqnarray}
\left\langle O_s\left| \psi_\uparrow^{\dag}({\bf R}-\frac{{\bf r}}{2}) \psi_\uparrow({\bf R}+\frac{{\bf r}}{2}) \right|I_s\right\rangle_{d}
&=& \int \frac{d^{3}{\bf p} dp_0}{(2\pi)^4} \frac{[-iT_{\mathbf{k},\mathbf{k}'}^{(s)}(k)]^2 i^3 e^{i{\bf p}\cdot{\bf r}}}{[p_0 - (-{\bf p})^2/(2m) + i0^+][k^2/m - p_0 - {\bf p}^2/(2m) + i0^+]^2} \nonumber \\
&=& \frac{im^2[T_{\mathbf{k},\mathbf{k}'}^{(s)}(k)]^2}{8\pi k} - \frac{r}{8\pi} m^2[T_{\mathbf{k},\mathbf{k}'}^{(s)}(k)]^2 + {\cal O}(r^2) + \cdot\cdot\cdot. \label{eq:OneBodyOperators}
\end{eqnarray}
\end{widetext}

\begin{figure}
\includegraphics[width=7cm]{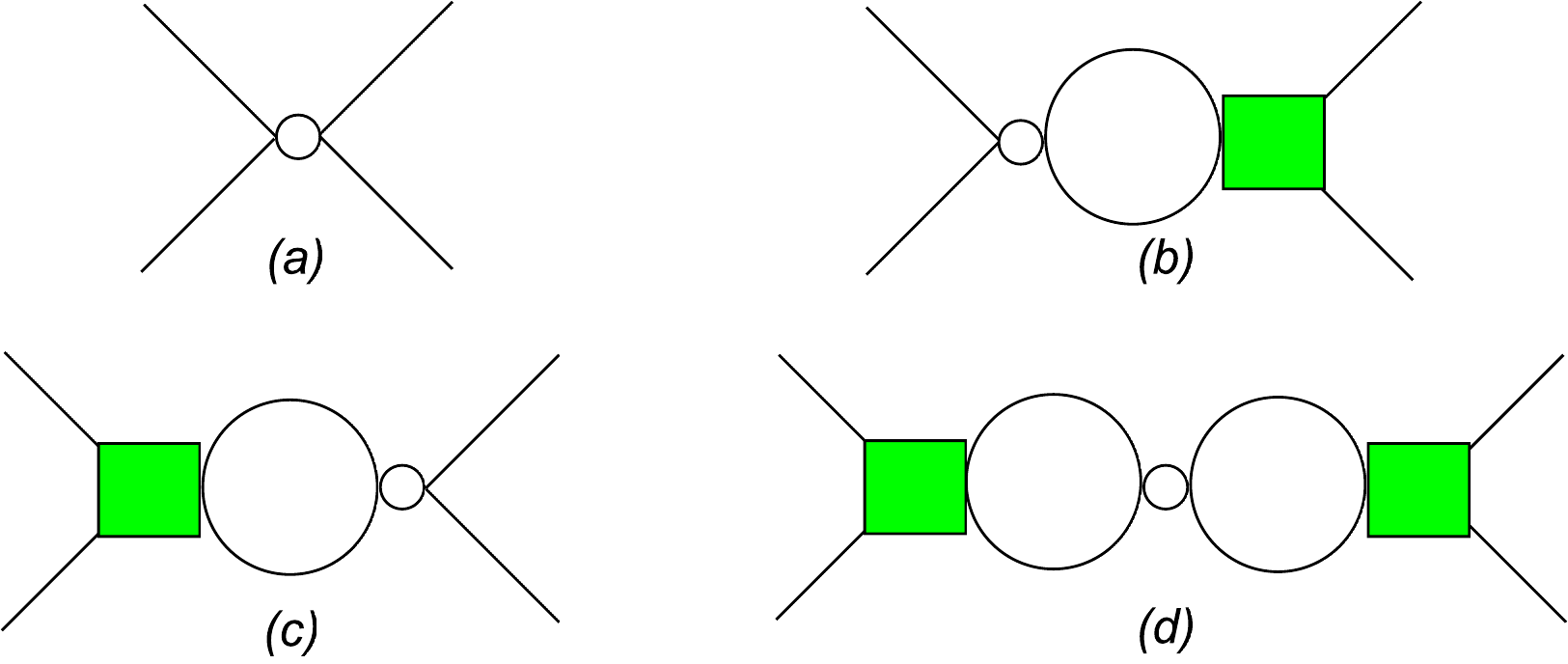}
\caption{(Color online) Diagrams for matrix elements of the
two-atom local operator $\psi^{\dag}_{\uparrow}({\bf R}) \psi^{\dag}_{\downarrow}({\bf R}) \psi_{\downarrow}({\bf R}) \psi_{\uparrow}({\bf R})$ and its derivatives. \label{fig:TwoBodyOperatorRs}}
\end{figure}

To match the nonanalytic terms in Eq.~(\ref{eq:OneBodyOperators}), we calculate the expectation values of the
two-atom operator $\langle O_s| \psi^{\dag}_{\uparrow}({\bf R}) \psi^{\dag}_{\downarrow}({\bf R}) \psi_{\downarrow}({\bf R}) \psi_{\uparrow}({\bf R}) |I_s \rangle$ as shown in Fig.~\ref{fig:TwoBodyOperatorRs}:
\begin{eqnarray}
&& \langle O_s| \psi^{\dag}_{\uparrow}({\bf R}) \psi^{\dag}_{\downarrow}({\bf R}) \psi_{\downarrow}({\bf R}) \psi_{\uparrow}({\bf R}) |I_s \rangle \nonumber \\
&=& \sum_{j=a,b,c,d} \langle O_s| \psi^{\dag}_{\uparrow}({\bf R}) \psi^{\dag}_{\downarrow}({\bf R}) \psi_{\downarrow}({\bf R}) \psi_{\uparrow}({\bf R}) |I_s \rangle_{j} \nonumber\\
&=& [ 1 -iT_{\mathbf{k},\mathbf{k}'}^{(s)}(k) \Pi_{s}(k) ]^2. \label{eq:TwoBodyOperatorsR1}
\end{eqnarray}

Substituting Eq.~(\ref{eq:Ts}) into (\ref{eq:TwoBodyOperatorsR1}), we have
\begin{eqnarray}\label{eq:TwoBodyOperatorsR2}
\langle O_s| \psi^{\dag}_{\uparrow}({\bf R}) \psi^{\dag}_{\downarrow}({\bf R}) \psi_{\downarrow}({\bf R}) \psi_{\uparrow}({\bf R}) |I_s \rangle = \frac{[T_{\mathbf{k},\mathbf{k}'}^{(s)}(k)]^2}{u_{s}^{2}}.
\end{eqnarray}

\subsection{$p$-wave channel}\label{4.2}

\begin{figure}
\includegraphics[width=7cm]{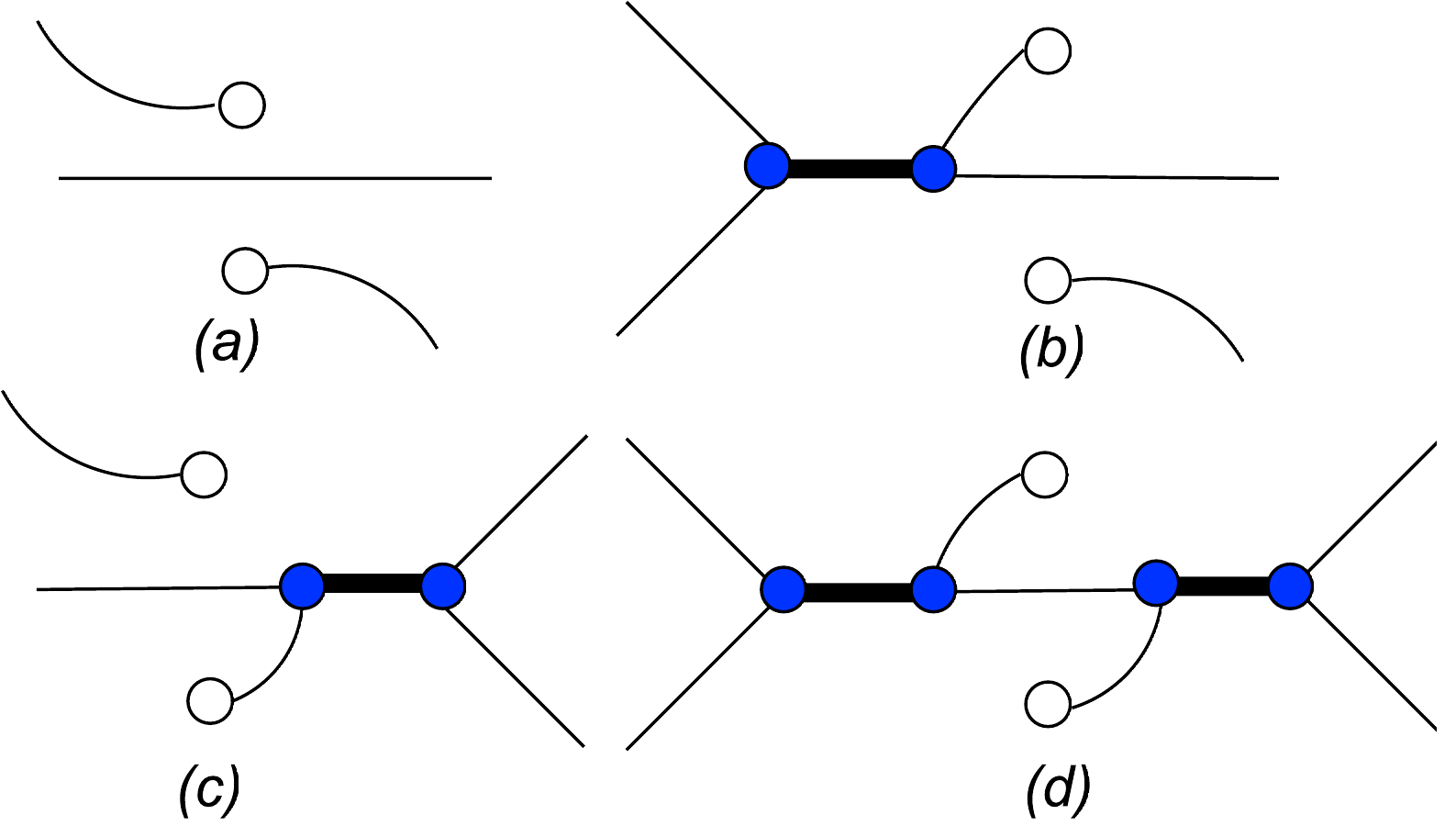}
\caption{(Color online) Diagrams for matrix elements of the operator $\psi_\uparrow^{\dag}({\bf R}-\frac{{\bf r}}{2}) \psi_\uparrow({\bf R}+\frac{{\bf r}}{2})$ in $p$-wave interacting channel. \label{fig:OneBodyOperatorp}}
\end{figure}

Similar to the case of $s$-wave interaction,
we can evaluate the diagram in Fig.~\ref{fig:OneBodyOperatorp}(d) as
\begin{widetext}
\begin{align}\label{eq:OneBodyOperatorp}
&~~\left\langle O_p\left| \psi_\uparrow^{\dag}({\bf R}-\frac{{\bf r}}{2}) \psi_\uparrow({\bf R}+\frac{{\bf r}}{2}) \right|I_p\right\rangle_{d} \nonumber \\
&= \sum_{m_I} k^{2}Y_{1,m_I}({\bf \hat{k}})Y^{*}_{1,m_I}({\bf \hat{k}}')D_{m_I}^{2}(k)\left(\frac{-ig_{m_I}}{\sqrt{2}}\right)^4 \nonumber \\
&~~ \times \int \frac{d^{3}{\bf p} dp_0}{(2\pi)^4} \frac{2 i^3 p^2 |Y_{1,m_I}({\bf \hat{p}})|^2 e^{i({\bf Q}/2+{\bf p})\cdot{\bf r}}}{[p_0 - ({\bf Q}/2-{\bf p})^2/(2m) + i0^+][q_0 - p_0 - ({\bf Q}/2+{\bf p})^2/(2m) + i0^+]^2} \nonumber \\
&\approx -\frac{m^2}{ 4\pi } \sum_{m_I} k^{2}Y_{1,m_I}({\bf \hat{k}})Y^{*}_{1,m_I}({\bf \hat{k}}')D_{m_I}^{2}(k)\left(\frac{-ig_{m_I}}{\sqrt{2}}\right)^4  \nonumber \\
&~~ \times \left[ \frac{1}{r} + \frac{i3 k }{2} - k^2 r - \frac{Q^2r}{24} + \left(i \frac{Q}{2} - \frac{3 k Q r }{4}\right)P_{1}(\hat{{\bf Q}}\cdot\hat{{\bf r}})  - \frac{Q^2r}{12}P_{2}(\hat{{\bf Q}}\cdot\hat{{\bf r}}) + {\cal O}(r^2) + \cdot\cdot\cdot \right],
\end{align}
\end{widetext}
where we average over the direction of ${\bf p}$ as an approximation, $q_{0} = k^2/m + Q^2/(4m)$ is the total incoming energy, $j_{l}(x)$ are the spherical Bessel functions, and $P_{l}(\hat{{\bf Q}}\cdot\hat{{\bf r}})$ are the Legendre polynomials.

\begin{figure}
\includegraphics[width=3cm]{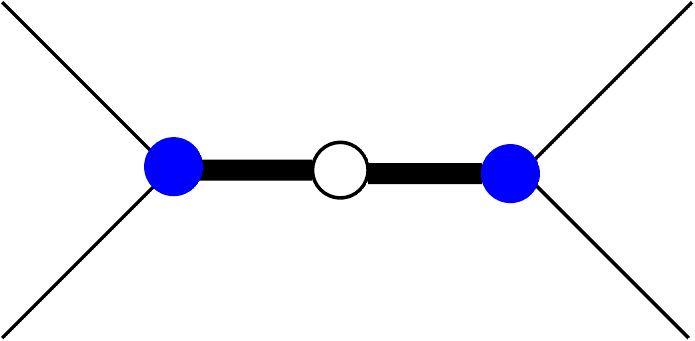}
\caption{(Color online) Diagram for matrix elements of the
one-molecule local operator $\varphi_{m_I}^\dagger({\bf R}) \varphi_{m_I}({\bf R})$ and its derivatives. \label{fig:TwoBodyOperatorRp}}
\end{figure}

To match the nonanalytic terms in Eq.~(\ref{eq:OneBodyOperatorp}), we calculate the expectation values of the
molecule operator $\varphi_{m_I}^\dagger({\bf R}) \varphi_{m_I}({\bf R})$ as shown in Fig.~\ref{fig:TwoBodyOperatorRp}:
\begin{eqnarray}
&&\langle O_p| \varphi_{m_I}^{\dag}({\bf R}) \varphi_{m_I}({\bf R}) |I_p\rangle \nonumber\\
&=& D_{m_I}^{2}(k)\left(\frac{-ig_{m_I}}{\sqrt{2}}\right)^{2} k^{2}Y_{1,m_I}({\bf \hat{k}})Y^{*}_{1,m_I}({\bf \hat{k}}'). \label{eq:TwoBodyOperatorpR1}
\end{eqnarray}

Therefore, we get
\begin{align}
&~~\left\langle O_p\left| \varphi_{m_I}^\dagger({\bf R})\left(i\partial_t+\frac{\nabla^2_{{\bf R}}}{4m}\right) \varphi_{m_I}({\bf R}) \right|I_p\right\rangle \nonumber\\
&= \frac{k^2}{m}D_{m_I}^{2}(k)\left(\frac{-ig_{m_I}}{\sqrt{2}}\right)^{2} k^{2}Y_{1,m_I}({\bf \hat{k}})Y^{*}_{1,m_I}({\bf \hat{k}}'), \label{eq:TwoBodyOperatorpR3}
\end{align}
\begin{align}
&~~\left\langle O_p\left| \varphi_{m_I}^\dagger({\bf R})\left(-i\nabla_{{\bf R}}\right) \varphi_{m_I}({\bf R}) \right|I_p\right\rangle \nonumber\\
&= {\bf Q}D_{m_I}^{2}(k)\left(\frac{-ig_{m_I}}{\sqrt{2}}\right)^{2} k^{2}Y_{1,m_I}({\bf \hat{k}})Y^{*}_{1,m_I}({\bf \hat{k}}'), \label{eq:TwoBodyOperatorpR4}
\end{align}
\begin{align}
&~~\left\langle O_p\left| \varphi_{m_I}^\dagger({\bf R})\left(-\frac{\nabla^2_{{\bf R}}}{4m}\right) \varphi_{m_I}({\bf R}) \right|I_p\right\rangle \nonumber\\
&= \frac{Q^2}{4m}D_{m_I}^{2}(k)\left(\frac{-ig_{m_I}}{\sqrt{2}}\right)^{2} k^{2}Y_{1,m_I}({\bf \hat{k}})Y^{*}_{1,m_I}({\bf \hat{k}}'). \label{eq:TwoBodyOperatorpR5}
\end{align}

\subsection{Coexistence of $s$- and $p$-wave channels}\label{4.3}

Matching Eq.~(\ref{eq:OneBodyOperators}) and Eq.~(\ref{eq:OneBodyOperatorp}) with Eq.~(\ref{eq:TwoBodyOperatorsR2}) and Eqs.~(\ref{eq:TwoBodyOperatorpR1})-(\ref{eq:TwoBodyOperatorpR5}), we get the momentum distribution $n_{\uparrow}(q)$ in the large $q$-limit ($n^{1/3}\ll q\ll 1/r_0$ with $n$ the total number density and $r_0$ the interaction range)
\begin{align}\label{eq:tail-up}
&n_{\uparrow}({\bf q}) =  \frac{C_a}{q^{4}V} + \frac{\sum_{m_I} C_{\upsilon,m_I}}{ q^2 V} - \frac{\sum_{m_I} \hat{{\bf q}}\cdot{\bf C}_{Q1,m_I}}{ q^3 V} \nonumber\\
&~~ + \frac{\sum_{m_I} [2C_{R,m_I} - C_{Q2,m_I} + 4C_{Q2,m_I}(\hat{{\bf q}}\cdot\hat{{\bf Q}})^2]}{ q^4 V},
\end{align}
where the corresponding contacts are defined as
\begin{align}
&C_{a} \equiv m^2 u_{s}^{2} \int d^{3}{\bf R} \langle \psi^{\dag}_{\uparrow}({\bf R}) \psi^{\dag}_{\downarrow}({\bf R}) \psi_{\downarrow}({\bf R}) \psi_{\uparrow}({\bf R}) \rangle, \label{ca}\\
&C_{\upsilon,m_I} \equiv m^2 g_{m_I}^2 \int d^{3}{\bf R} \langle \varphi_{m_I}^\dagger({\bf R}) \varphi_{m_I}({\bf R}) \rangle, \label{cvmI}\\
&C_{R,m_I} \equiv m^3 g_{m_I}^2  \nonumber\\
&~~\times\int d^{3}{\bf R} \left\langle \varphi_{m_I}^\dagger({\bf R})\left(i\partial_t+\frac{\nabla^2_{{\bf R}}}{4m}\right) \varphi_{m_I}({\bf R}) \right\rangle, \label{cRmI}\\
&{\bf C}_{Q1,m_I} \equiv m^2 g_{m_I}^2 \int d^{3}{\bf R} \left\langle \varphi_{m_I}^\dagger({\bf R})\left(-i\nabla_{{\bf R}}\right) \varphi_{m_I}({\bf R}) \right\rangle, \label{cQ1mI}\\
&C_{Q2,m_I} \equiv m^3 g_{m_I}^2 \int d^{3}{\bf R} \left\langle \varphi_{m_I}^\dagger({\bf R})\left(-\frac{\nabla^2_{{\bf R}}}{4m}\right) \varphi_{m_I}({\bf R}) \right\rangle. \label{cQ2mI}
\end{align}
Notice that the distribution of ${\bf Q}$ here is anisotropic. Therefore, we find that ${\bf C}_{Q1,m_I}$ is anisotropic and the $q^{-3}$ tail and part of the $q^{-4}$ tail of the momentum distribution Eq.~(\ref{eq:tail-up}) show anisotropic behaviors of center-of-mass momentum ${\bf Q}$.
Especially in the previous studies, it has been shown that the contacts of a similar nature to ${\bf C}_{Q1,m_I}$ in $q^{-3}$ tail and $C_{Q2,m_I}$ in $q^{-4}$ tail also exist for one-dimensional $p$-wave Fermi gases~\cite{Cui20162}.

Note that, if we consider finite total momentum in the $s$-wave case, the center-of-mass-momentum-related contacts will appear in the $q^{-5}$ and $q^{-6}$ tails of the high-momentum distribution~\cite{Yi2018}, where $q$ is the relative momentum. However, we calculate only the high momentum tails up to $q^{-4}$, so that the $q^{-5}$ and $q^{-6}$ tails are not included here. Therefore, we consider only zero total momentum in the $s$-wave case.

We emphasize that, in our system, the center-of-mass momentum induces the anisotropic behavior of the contact. Here, only ${\bf C}_{Q1,m_I}$ in $q^{-3}$ tail and $C_{Q2,m_I}$ in $q^{-4}$ tail are both center-of-mass-momentum and laser dependent, but only ${\bf C}_{Q1,m_I}$ is anisotropic. The other three contacts are laser dependent but not center-of-mass-momentum dependent.

As the adiabatic relations shown in the next section, $C_{a}$, $C_{\upsilon,m_I}$ and $C_{R,m_I}$ are associated to the inverse of $s$-wave scattering length, the inverse of $p$-wave scattering volume, and the inverse of $p$-wave effective range.
The last two contacts ${\bf C}_{Q1,m_I}$ and $C_{Q2,m_I}$ are related to the velocity and the kinetic energy of the closed-channel molecules, respectively.

A simple physical picture to describe the anisotropic behavior of the contact is as follows. The contact is a many-body physical quantity which bridges the few- and many-body physics. Therefore, the anisotropic behavior of the contact is dependent both on the center-of-mass momentum and the anisotropy of the many-body wave function of the system. For example, a finite anisotropic contact can be probed in the Fulde-Ferrell state which supports a finite-momentum two-body bound state and pairing superfluidity~\cite{Yi2018}. Specifically, the anisotropic contact induced by the center-of-mass momentum appears only in the subleading tails of the pure $s$- and pure $p$-wave high-momentum distributions, respectively.

We should clarify that, in this system, the center-of-mass momentum is a good quantum number and the Galilean invariance is not broken.
Therefore, the high-momentum tail in the presence of the finite center-of-mass momentum can be obtained by doing a frame transformation from the center-of-mass frame.
For concreteness, we explicitly assume that the system has a distribution of the center-of-mass momentum, which accounts for the anisotropy in the high-momentum tail.

For simplicity, in our model, we assume that the $p$-wave interaction exists only between two spin-up fermions, and there is no interaction between two spin-down fermions.
Therefore, the momentum distribution for the spin-down fermions has only the $s$-wave contact.
\begin{align}\label{eq:tail-down}
n_{\downarrow}(\mathbf q) = \frac{C_a}{Vq^{4}}.
\end{align}

\section{Universal relations}\label{5}

In this section, we derive the corresponding universal relations.

\subsection{High-frequency radio-frequency spectroscopy}\label{5.1}

The rf spectroscopy can be used as an important experimental tool to detect the contacts~\cite{Yu2015,rf2010s-wave1,rf2010s-wave2,Hofmann2011,Thompson2010}.
The high-frequency tails of the rf spectroscopy are governed by contacts.
The rf with frequency $\omega$ is applied to transfers fermions from the internal spin state $|\sigma\rangle$ ($\sigma=\uparrow,\downarrow$) into a third spin state $|3\rangle$.
The resultant number of the atoms transferred to state $|3\rangle$ is proportional to
the transition rate, which is given by~\cite{rf2010s-wave2,Hofmann2011}
\begin{align}\label{eq:rf}
\Gamma_{rf,\sigma}(\omega) &= \Omega_{rf}^{2} \text{Im}~i\int d^{3}{\bf R} \int dt e^{i\omega t} \int d^{3}{\bf r} \nonumber \\
&~~ \times\left\langle {\cal T}{\cal O}^{\dag}_{\sigma3}({\bf R}+\frac{{\bf r}}{2},t){\cal O}_{\sigma3}({\bf R}-\frac{{\bf r}}{2},0) \right\rangle,
\end{align} where $\Omega_{rf}$ is the rf Rabi frequency determined by the strength of the rf
signal, ${\cal O}_{\sigma3}({\bf r},t) \equiv \psi_{3}^{\dag}({\bf r},t)\psi_{\sigma}({\bf r},t)$, and ${\cal T}$ is the time ordering operator.

\begin{figure}
\includegraphics[width=7cm]{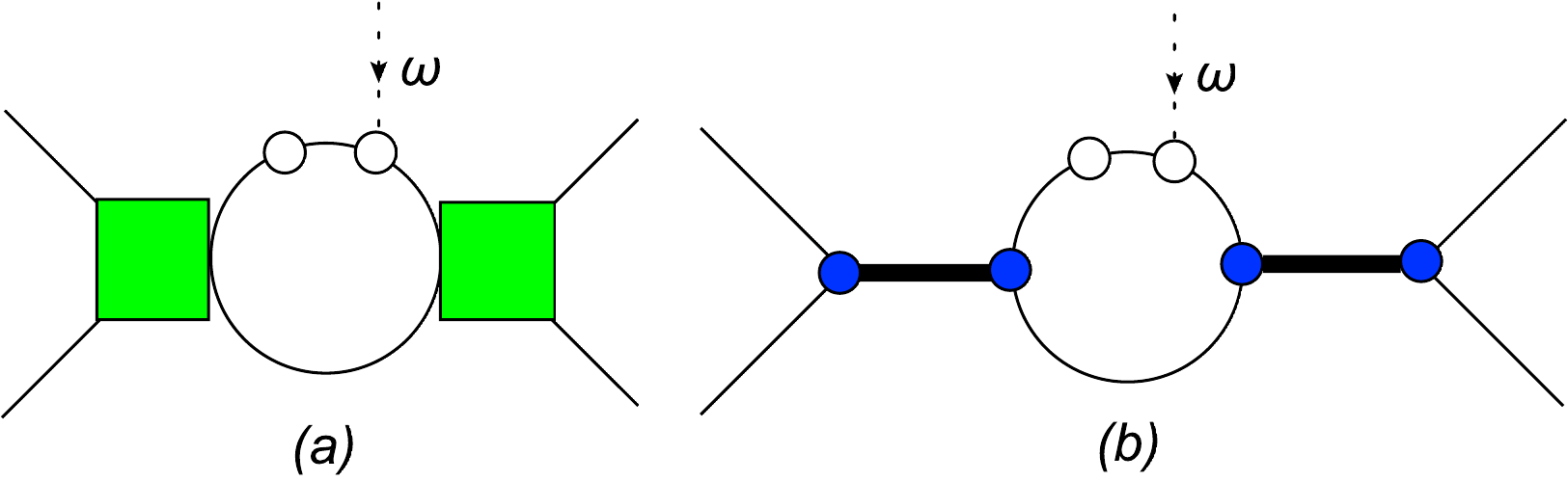}
\caption{(Color online) Diagrams for the matrix element of $\int dt e^{i\omega t}\int d^{3}{\bf r} {\cal T}{\cal O}_{\sigma3}^{\dag}({\bf R}+\frac{{\bf r}}{2},t){\cal O}_{\sigma3}({\bf R}-\frac{{\bf r}}{2},0)$ ($\sigma=\uparrow,\downarrow$). (a) the $s$-wave case and (b) the $p$-wave case. \label{fig:rf}}
\end{figure}

We can evaluate the diagram in Figs.~\ref{fig:rf}(a) and \ref{fig:rf}(b) as
\begin{widetext}
\begin{align}
&~~\int dt e^{i\omega t}\int d^{3}{\bf r} \left\langle O_s\left| {\cal T}{\cal O}^{\dag}_{\sigma3}({\bf R}+\frac{{\bf r}}{2},t){\cal O}_{\sigma3}({\bf R}-\frac{{\bf r}}{2},0) \right|I_s\right\rangle \nonumber \\
&= \int \frac{d^{3}{\bf p} dp_0}{(2\pi)^4} \frac{ i^4 [-iT_{\mathbf{k},\mathbf{k}'}^{(s)}(k)]^2 }{[p_0 - ({\bf Q}/2-{\bf p})^2/(2m) + i0^+][q_0 - p_0 - ({\bf Q}/2+{\bf p})^2/(2m) + i0^+]^2 [q_0 - p_0 + \omega - ({\bf Q}/2+{\bf p})^2/(2m) + i0^+]} \nonumber \\
&= \frac{m^3[T_{\mathbf{k},\mathbf{k}'}^{(s)}(k)]^2}{4\pi} \left[ \frac{1}{(m\omega)^{3/2}} + \frac{k^2}{2(m\omega)^{3/2}} - \frac{1}{2|k|m\omega} - \frac{|k|}{(m\omega)^2} + {\cal O}((m\omega)^{-5/2}) + \cdot\cdot\cdot \right], \label{eq:rf-s}
\end{align}
\end{widetext}
\begin{widetext}
\begin{align}
&~~\int dt e^{i\omega t}\int d^{3}{\bf r} \left\langle O_p\left| {\cal T}{\cal O}^{\dag}_{\uparrow3}({\bf R}+\frac{{\bf r}}{2},t){\cal O}_{\uparrow3}({\bf R}-\frac{{\bf r}}{2},0) \right|I_p\right\rangle \nonumber \\
&= \sum_{m_I} k^{2}Y_{1,m_I}({\bf \hat{k}})Y^{*}_{1,m_I}({\bf \hat{k}}')D_{m_I}^{2}(k)\left(\frac{-ig_{m_I}}{\sqrt{2}}\right)^4 \nonumber \\
&~~ \times \int \frac{d^{3}{\bf p} dp_0}{(2\pi)^4} \frac{ 2i^4 p^2 |Y_{1,m_I}({\bf \hat{p}})|^2 }{[p_0 - ({\bf Q}/2-{\bf p})^2/(2m) + i0^+][q_0 - p_0 - ({\bf Q}/2+{\bf p})^2/(2m) + i0^+]^2 [q_0 - p_0 + \omega - ({\bf Q}/2+{\bf p})^2/(2m) + i0^+]} \nonumber \\
&= \sum_{m_I} k^{2}Y_{1,m_I}({\bf \hat{k}})Y^{*}_{1,m_I}({\bf \hat{k}}')D_{m_I}^{2}(k)\left(\frac{-ig_{m_I}}{\sqrt{2}}\right)^2 \frac{m^{3} g_{m_I}^{2}}{16\pi^2} \left[ \frac{1}{(m\omega)^{1/2}} + \frac{3k^2}{2(m\omega)^{3/2}} - \frac{3|k|}{2m\omega} - \frac{|k|^3}{(m\omega)^2} + {\cal O}((m\omega)^{-5/2}) + \cdot\cdot\cdot \right].\label{eq:rf-p}
\end{align}
\end{widetext}

Matching Eq.~(\ref{eq:rf-s}) and Eq.~(\ref{eq:rf-p}) with Eq.~(\ref{eq:TwoBodyOperatorsR2}) and Eqs.~(\ref{eq:TwoBodyOperatorpR1})-(\ref{eq:TwoBodyOperatorpR5}), we have the rf transfer rate from Eq.~(\ref{eq:rf}) in high-frequency limit ($1/(mR^{2}_{m_I})\gg\omega\gg E_{F}$ with Fermi energy $E_{F}=k^{2}_{F}/(2m)$ and Fermi wave vector $k_F$):
\begin{align}
\Gamma_{rf,\uparrow}(\omega) &= \frac{m\Omega_{rf}^{2}}{4\pi} \left\{ \frac{C_{a}}{(m\omega)^{3/2}} \right. \nonumber\\
&\left. ~~ + \sum_{m_I} \left[ \frac{C_{\upsilon,m_I}}{(m\omega)^{1/2}} + \frac{3C_{R,m_I}}{2(m\omega)^{3/2}} \right] \right\},\\
\Gamma_{rf,\downarrow}(\omega) &= \frac{m\Omega_{rf}^{2}}{4\pi} \frac{C_{a}}{(m\omega)^{3/2}}.
\end{align}

\subsection{Adiabatic relations}\label{5.2}

With the Hellmann-Feynman theorem and Eqs.~(\ref{ca}), (\ref{cvmI}), and (\ref{cRmI}), we obtain the adiabatic relations
\begin{align}
& \frac{\partial E}{\partial a_{s}^{-1}} = -\int d^{3}{\bf R} \left\langle \frac{\partial {\cal L}}{\partial a_{s}^{-1}} \right\rangle = -\frac{C_{a}}{4\pi m}, \label{eq:caE}\\
& \frac{\partial E}{\partial \upsilon_{m_I}^{-1}} = - \int d^{3}{\bf R} \left\langle \frac{\partial {\cal L}}{\partial \upsilon_{m_I}^{-1}} \right\rangle = -\frac{C_{\upsilon,m_I}}{4\pi m}, \label{eq:cvE}\\
& \frac{\partial E}{\partial R_{m_I}^{-1}} =  - \int d^{3}{\bf R} \left\langle \frac{\partial {\cal L}}{\partial R_{m_I}^{-1}} \right\rangle = -\frac{C_{R,m_I}}{4\pi m}, \label{eq:cRE}
\end{align} where $E$ is the total energy of the many-body system, and we have used the following relations:
\begin{align}
&\left\langle \frac{\partial {\cal L}}{\partial a_{s}^{-1}} \right\rangle
= \left\langle \frac{\partial {\cal L}}{\partial u_{s}} \right\rangle \frac{\partial u_{s}}{\partial a_{s}^{-1}} \nonumber\\
&= \frac{mu_{s}^2}{4\pi} \langle \psi^{\dag}_{\uparrow}({\bf R}) \psi^{\dag}_{\downarrow}({\bf R}) \psi_{\downarrow}({\bf R}) \psi_{\uparrow}({\bf R}) \rangle, \label{ca0}\\
&\left\langle \frac{\partial {\cal L}}{\partial \upsilon_{m_I}^{-1}} \right\rangle
= \left\langle \frac{\partial {\cal L}}{\partial \nu_{m_I}} \right\rangle \frac{\partial \nu_{m_I}}{\partial \upsilon_{m_I}^{-1}} \nonumber\\
&= \frac{m g_{m_I}^{2}}{16\pi^2} \left\langle \varphi_{m_I}^\dagger({\bf R}) \varphi_{m_I}({\bf R}) \right\rangle, \label{cv0}\\
&\left\langle \frac{\partial {\cal L}}{\partial R_{m_I}^{-1}} \right\rangle
= \left\langle \frac{\partial {\cal L}}{\partial g_{m_I}} \right\rangle \frac{\partial g_{m_I}}{\partial R_{m_I}^{-1}} + \left\langle \frac{\partial {\cal L}}{\partial \nu_{m_I}} \right\rangle \frac{\partial \nu_{m_I}}{\partial g_{m_I}} \frac{\partial g_{m_I}}{\partial R_{m_I}^{-1}} \nonumber\\
&= \frac{m g_{m_I}^{2}}{16\pi^2} \left\langle \varphi_{m_I}^\dagger({\bf R})\left(i\partial_t+\frac{\nabla^2_{{\bf R}}}{4m}\right) \varphi_{m_I}({\bf R}) \right\rangle. \label{cR0}
\end{align}

\subsection{Pressure relation}\label{5.3}

For a uniform gas, the pressure relation can be derived following the expression of the Helmholtz free energy density ${\cal F}=F/V$ which can be expressed in terms of~\cite{Tan20083,Zhang2009,Yi2018,Zwerger2011,Yu2015,Cui20161,Cui20162,Platter2016}
%
%
\begin{align}
5{\cal F}&=
\sum_{m_I} \left(2T \frac{\partial}{\partial T} + 3n_{\downarrow}\frac{\partial}{\partial n_{\downarrow}} + 3n_{\uparrow}\frac{\partial}{\partial n_{\uparrow}} \right. \nonumber\\
&\left.~~ - a_{s}\frac{\partial}{\partial a_{s}} - 3\upsilon_{m_I}\frac{\partial}{\partial \upsilon_{m_I}} + R_{m_I}\frac{\partial}{\partial R_{m_I}}\right){\cal F}. \label{eq:F3}
\end{align}

Using the thermodynamical relations and the adiabatic relations (\ref{eq:caE})-(\ref{eq:cRE}), we can get the pressure relation as
\begin{eqnarray}
{\cal P} &=& \frac{2}{3}{\cal E} + \frac{C_a}{12\pi ma_{s}V} \nonumber\\
&& + \sum_{m_I} \left( \frac{C_{\upsilon,m_I}}{4\pi m\upsilon_{m_I}V} - \frac{C_{R,m_I}}{12\pi m R_{m_I}V} \right), \label{eq:pressure}
\end{eqnarray}
where $\cal P$ is the pressure density and $\cal E$ is the energy density.

\subsection{Virial theorem}\label{5.4}

For an atomic gas in a harmonic potential $V_T=\sum_j m\omega^2 {\bf r}_j^2/2$, the total energy can be expressed in terms of~\cite{Tan20083,Zhang2009,Yi2018,Zwerger2011,Yu2015,Cui20161,Cui20162,Platter2016}
\begin{align}
E &= \sum_{m_I} \left(\omega \frac{\partial}{\partial \omega} - \frac{1}{2}a_{s} \frac{\partial}{\partial a_{s}} \right. \nonumber\\
&\left.~~ - \frac{3}{2}\upsilon_{m_I} \frac{\partial}{\partial \upsilon_{m_I}} + \frac{1}{2}R_{m_I}\frac{\partial}{\partial R_{m_I}}\right)E,
\end{align}
which, together with the Hellmann-Feynman theorem and the adiabatic relations (\ref{eq:caE})-(\ref{eq:cRE}), gives
\begin{equation}
E = 2 \langle V_T\rangle - \frac{C_a}{8\pi m a_{s}} - \sum_{m_I} \left( \frac{3C_{\upsilon,m_I}}{8\pi m \upsilon_{m_I}} - \frac{C_{R,m_I}}{8\pi m R_{m_I}} \right).  \label{eq:virial}
\end{equation}

\section{Quantum virial expansion}\label{6}

The idea of the quantum virial expansion is to expand the thermodynamic quantities in powers of the fugacity $z_{\sigma}=e^{\beta\mu_{\sigma}}$, where $\beta=1/T$ and $T$ is the temperature.

In order to investigate the experimental detectable many-body physics of the above system, we calculate the normal-state contacts and spectral function of the system by using the quantum virial expansions~\cite{Kaplan2011,Leyronas2011,Leyronas2015,Barth2015,Cui20171,Cui20172,diagram2D2013,diagram2D2014,diagram3D2018,Liu2009,Liu20101,Liu20102,Liu20103,Liureview2013,PengPLA2011,PengPRA2011,HoNarrow2012,PengNarrow2014,Pengvirial2018,Ho2004s,Ho2004p,Qin2017}.

\subsection{Thermodynamic potential}\label{6.1}

To further calculate the normal-state contacts of the system, we will first evaluate the thermodynamic potential as follows.

In our model, two fermionic atoms with spin-$\uparrow$ species interact with each other by exchanging a bosonic molecule as shown in Fig.~\ref{fig:Tp}.
Therefore, we can write the grand thermodynamic potential of a strongly interacting Fermi gas as (up to the second order)~\cite{Liu2009,Liu20101,Liu20102,Liu20103,Liureview2013}
\begin{eqnarray}\label{omegah}
\Omega
&=& - T\frac{V}{\lambda^{3}}\left[ f_{5/2}(z_{\uparrow}) + f_{5/2}(z_{\downarrow}) \right. \nonumber\\
&& \left. + 2z_{\uparrow}z_{\downarrow}\Delta b_{2,s}
+ 2z_{\uparrow}^2 \Delta b_{2,p} \right],
\end{eqnarray}
where $\Delta b_{2,s}$ is the second virial coefficient which includes the $s$-wave two-body interaction shown in Fig.~\ref{fig:Ts}, $\Delta b_{2,p}$ is the second virial coefficient which includes the physical $p$-wave two-body interaction shown in Fig.~\ref{fig:Tp}, $\lambda\equiv\sqrt{2\pi/(mT)}$ is the thermal de Broglie wavelength, and $f_{\upsilon}(z_{\sigma})=[1/\Gamma(\upsilon)]\int_{0}^{\infty}x^{\upsilon-1}dx/(z_{\sigma}^{-1}e^{x} + 1)$ is the Fermi-Dirac integral with the gamma function $\Gamma(\upsilon)$~\cite{Pathria1996}.

\subsection{Normal-state contacts}\label{6.3}

According to the adiabatic relations Eqs.~(\ref{eq:caE})-(\ref{eq:cRE}), the contacts can also be expressed in term of the grand thermodynamic potential $\Omega$~\cite{Liureview2013,Ohashi2018}:
\begin{eqnarray}
C_{a} &=& -4\pi m  \left(\frac{\partial \Omega}{\partial a_{s}^{-1}}\right)_{T,V,\mu_{\uparrow},\mu_{\downarrow}}, \label{Ca}\\
C_{\upsilon,m_I} &=& -4\pi m  \left(\frac{\partial \Omega}{\partial \upsilon_{m_I}^{-1}}\right)_{T,V,\mu_{\uparrow},\mu_{\downarrow}}, \label{Cv}\\
C_{R,m_I} &=& -4\pi m \left(\frac{\partial \Omega}{\partial R_{m_I}^{-1}}\right)_{T,V,\mu_{\uparrow},\mu_{\downarrow}}. \label{CR}
\end{eqnarray}

\subsection{Self-energy}\label{6.4}

In order to calculate the normal-state self-energy, we first expand the noninteracting fermionic Green's function in powers of the fugacity $z_{\sigma}=e^{\beta\mu_{\sigma}}$~\cite{Leyronas2011,Leyronas2015,Barth2015,Cui20171,Cui20172}
\begin{align}\label{eq:Green function}
G_{\sigma}^{(0)}({\bf k},\tau) = e^{\mu_{\sigma}\tau} \sum_{n\geqslant0} G_{\sigma}^{(0,n)}({\bf k},\tau) z_{\sigma}^{n},
\end{align} where
$G_{\sigma}^{(0,0)}({\bf k},\tau) = -\Theta(\tau) e^{\mu_{\sigma}\tau}$,
$G_{\sigma}^{(0,n)}({\bf k},\tau) = (-1)^{n-1} e^{-\varepsilon_{\bf k}\tau} e^{-\varepsilon_{\bf k}n\beta}$ with $n\geqslant1$,
$\Theta(\tau)$ is the Heaviside function, $\varepsilon_{\bf k}=k^2/(2m)$, and $\tau$ is the imaginary time.

\begin{figure}
\includegraphics[width=7cm]{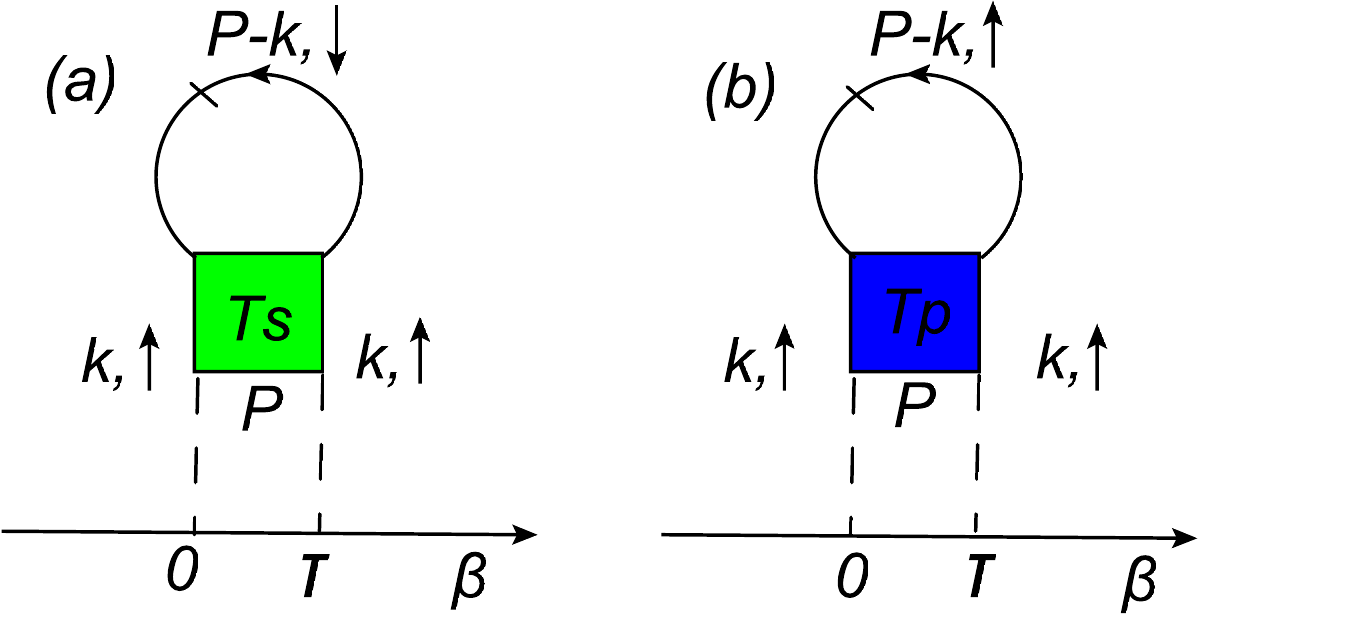}
\caption{(Color online) Lowest order diagram for the self-energy for (a) the $s$-wave interaction and (b) the $p$-wave interaction, respectively.
The fermion propagator line with $n$ vertical dashes denotes the $n$th-order
contribution $G_{\sigma}^{(0,n)}$ in Eq.~(\ref{eq:Green function}).
Unlike the Feynman diagrams in the above figures, here, the green square with $T$ inside represents the $T_s$ matrix: $T_{\mathbf{p},\mathbf{p}'}^{(s)}$ and the blue square with $T$ inside represents the $T_p$ matrix: $T_{\mathbf{p},\mathbf{p}'}^{(m_I)}$.
} \label{fig:self-energy}
\end{figure}

According to the Feynman diagram shown in Fig.~\ref{fig:self-energy}, the lowest order of the self-energies for the $s$- and $p$-wave interactions are, respectively,
\begin{align}
\Sigma^{(s)}_{\sigma}({\bf k},\tau) &= \int\frac{d^3{\bf P}}{(2\pi)^3} e^{\mu_{\sigma}\tau} [G_{\sigma}^{(0,1)}({\bf k},-\tau) z_{\sigma}] T_{\mathbf{p},\mathbf{p}'}^{(s)}({\bf P},\tau), \\
\Sigma^{(p)}_{m_I}({\bf k},\tau) &= \int\frac{d^3{\bf P}}{(2\pi)^3} e^{\mu_{\uparrow}\tau} [G_{\uparrow}^{(0,1)}({\bf k},-\tau) z_{\uparrow}] T_{\mathbf{p},\mathbf{p}'}^{(m_I)}({\bf P},\tau),
\end{align} where
\begin{align}
T_{\mathbf{p},\mathbf{p}'}^{(s)}({\bf P},\tau)
&= e^{-\frac{P^2}{4m}\tau} \int_{\gamma-i\infty}^{\gamma+i\infty}\frac{dz}{2\pi i} e^{- \tau z}\frac{4\pi/m}{a_{s}^{-1} - \sqrt{-mz}}, \\
T_{\mathbf{p},\mathbf{p}'}^{(m_I)}({\bf P},\tau)
&= e^{-\frac{P^2}{4m}\tau} \int_{\gamma-i\infty}^{\gamma+i\infty} \frac{dz}{2\pi i} e^{-\tau z} \nonumber\\
&~~\times\frac{16\pi^2 zY_{1,m_I}({\bf \hat{p}})Y^{*}_{1,m_I}({\bf \hat{p}}')}{ \upsilon_{m_I}^{-1} + R_{m_I}^{-1}mz - mz\sqrt{-mz} },
\end{align}
with the complex $z=\text{Re}(z)+i0^+$.

Therefore, we derive the retarded self-energy for spin $\sigma$ as
\begin{align}
\Sigma_{\uparrow}^{(R)}({\bf k},E_{\uparrow}) &= \Sigma^{(s)}_{\uparrow}({\bf k},E_{\uparrow}) + \sum_{m_I}\Sigma^{(p)}_{m_I}({\bf k},E_{\uparrow}) \nonumber\\
&= z_{\uparrow}[F^{(s)}({\bf k},E_{\uparrow})+F^{(p)}({\bf k},E_{\uparrow})] \nonumber\\
&~~ + z_{\uparrow}^{2}[H^{(s)}({\bf k},E_{\uparrow})+H^{(p)}({\bf k},E_{\uparrow})] , \label{eq:self-energy1}\\
\Sigma_{\downarrow}^{(R)}({\bf k},E_{\downarrow}) &= \Sigma^{(s)}_{\downarrow}({\bf k},E_{\downarrow}) \nonumber\\
&= z_{\downarrow}F^{(s)}({\bf k},E_{\downarrow}) + z_{\downarrow}^{2}H^{(s)}({\bf k},E_{\downarrow}) ,\label{eq:self-energy2}
\end{align} where $E_{\sigma}=\omega+\mu_{\sigma}+i0^+$, $F^{(s)}({\bf k},E_{\sigma})$, $F^{(p)}({\bf k},E_{\sigma})$, $H^{(s)}({\bf k},E_{\sigma})$, and $H^{(p)}({\bf k},E_{\sigma})$ are given in the Appendix.

\subsection{Normal-state spectral function}\label{6.5}

We calculate the spectral function as~\cite{Leyronas2015,Barth2015,Cui20171,Cui20172}
\begin{align}
A_{\sigma}({\bf k},E_{\sigma}) = - \frac{1}{\pi}\text{Im}[G_{\sigma}^{(R)}({\bf k},E_{\sigma})],
\end{align} where the retarded Green's function is given by
\begin{align}
G_{\sigma}^{(R)}({\bf k},E_{\sigma}) = \frac{1}{E_{\sigma}-k^2/(2m)-\Sigma_{\sigma}^{(R)}({\bf k},E_{\sigma})}.
\end{align}

\section{Numerical results}\label{7}

\begin{figure*}
\includegraphics[width=6cm]{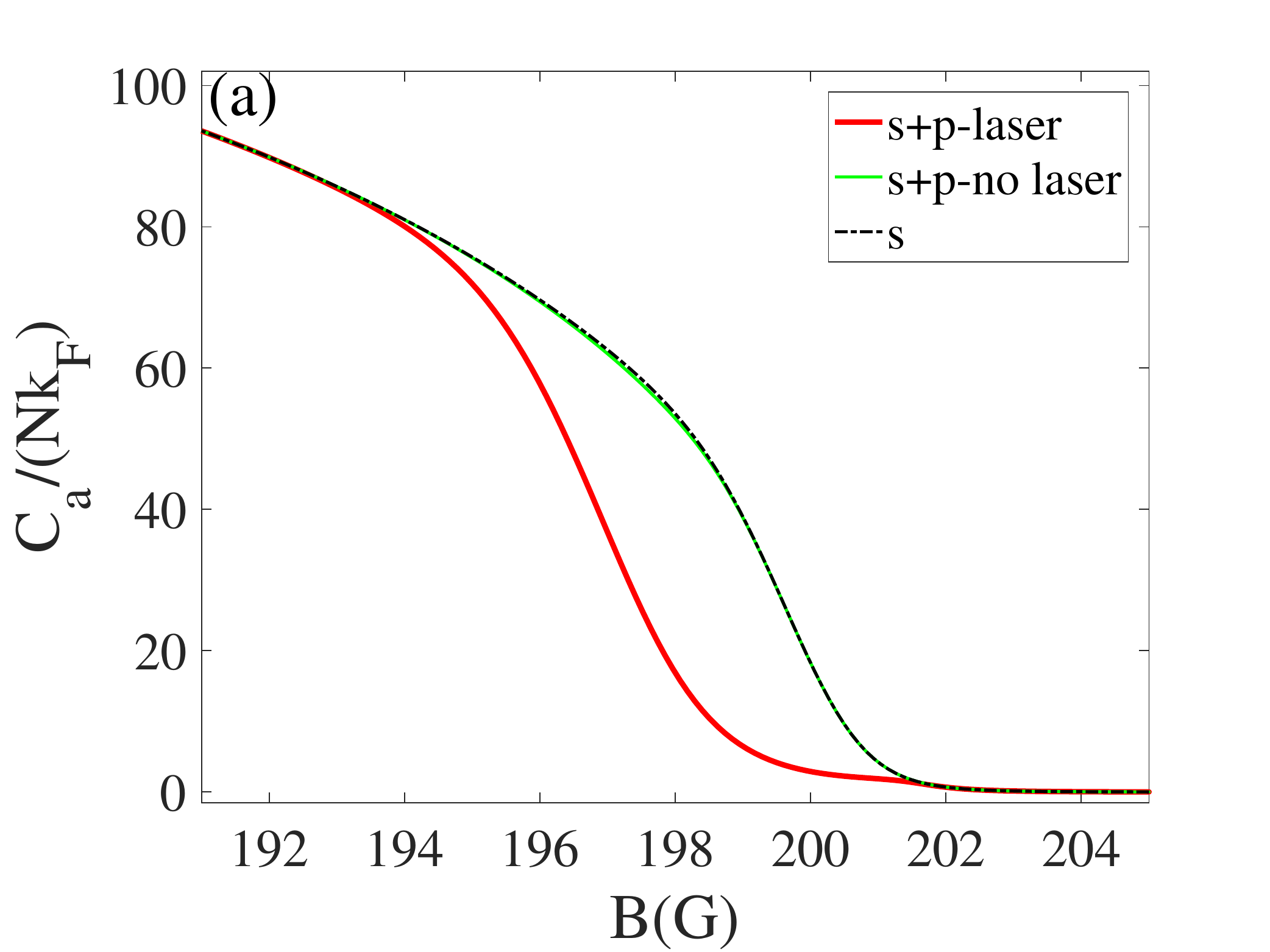}\includegraphics[width=6cm]{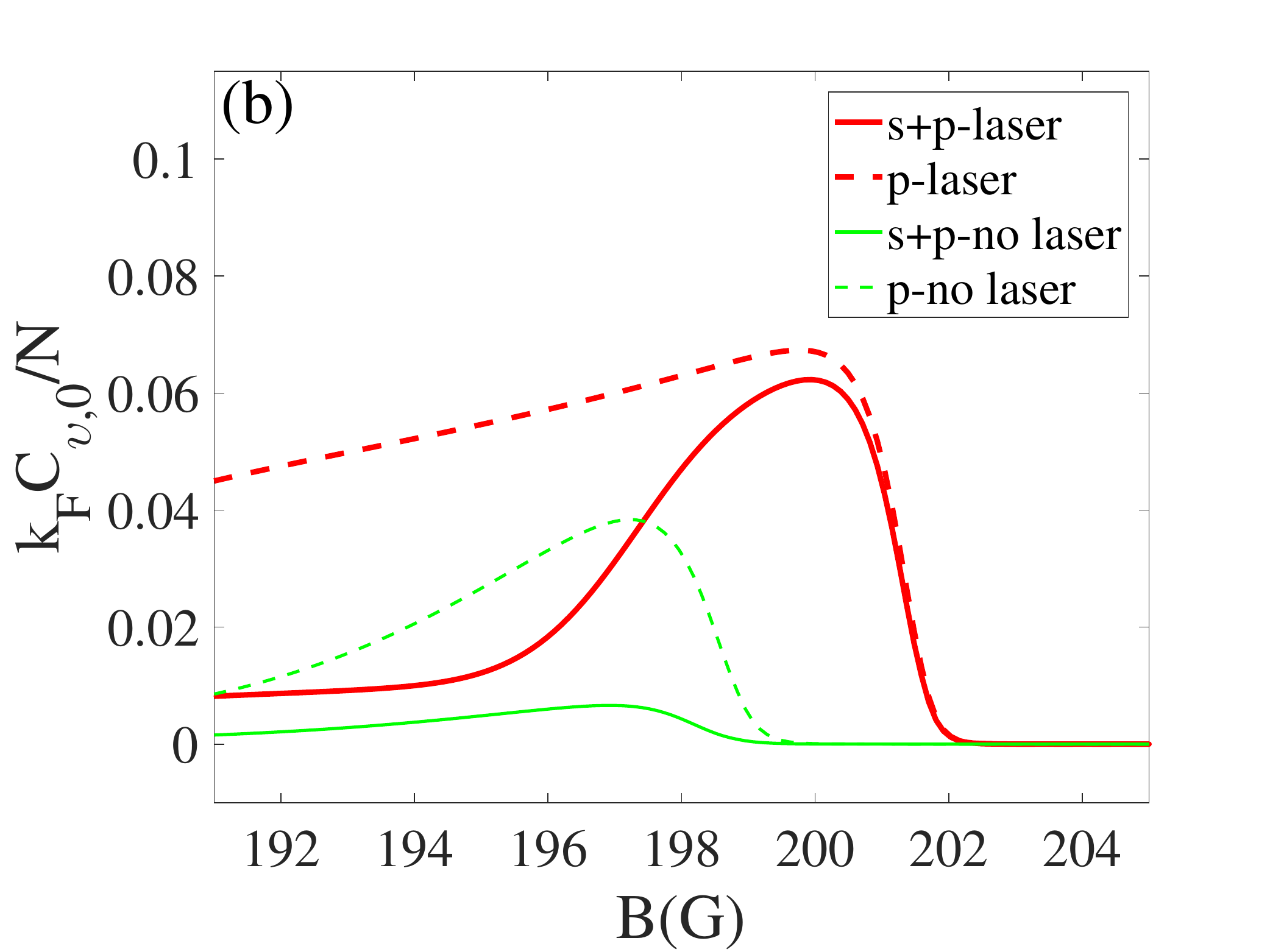}\includegraphics[width=6cm]{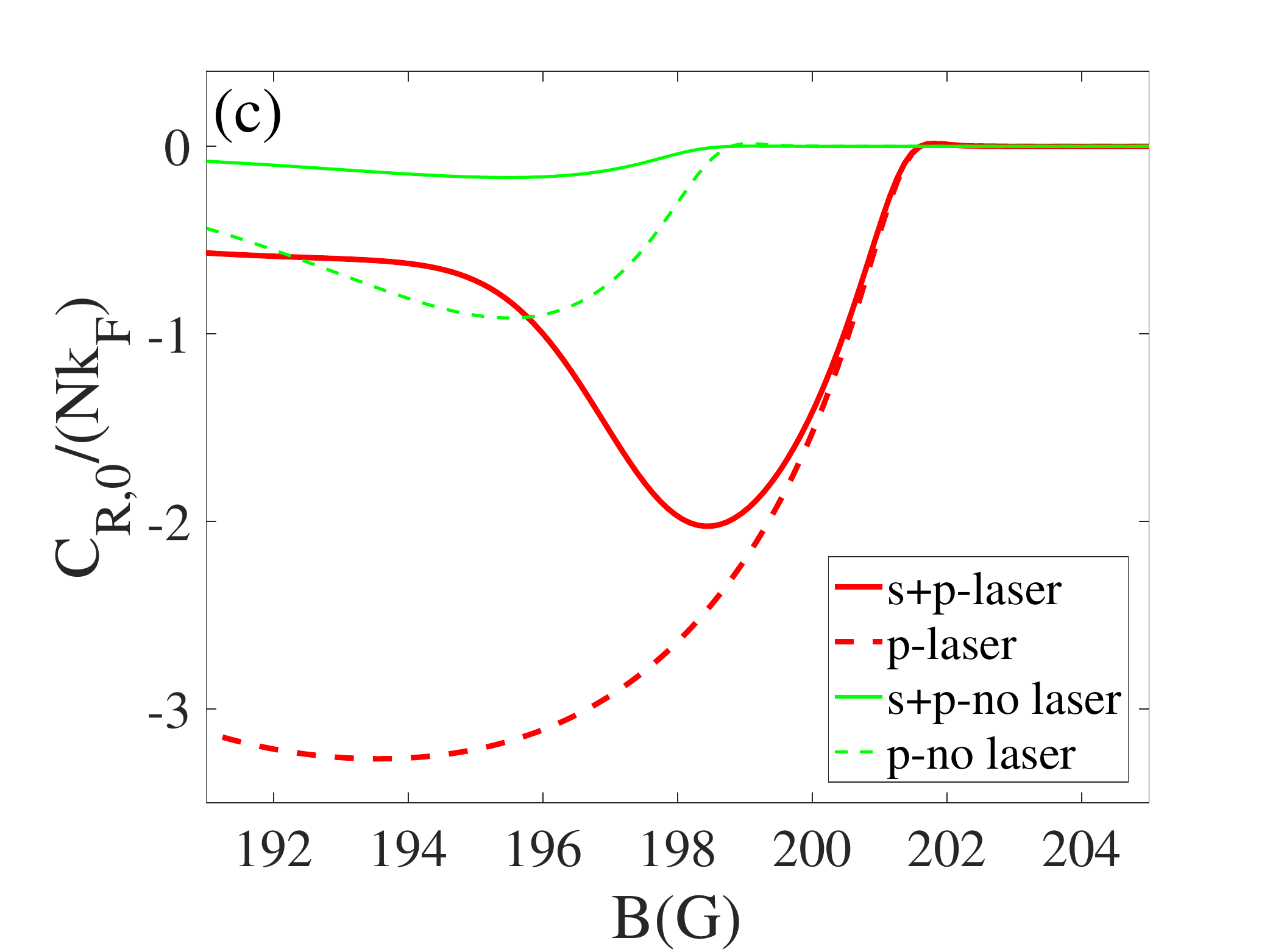}
\caption{(Color online) Contacts $C_{a}$, $C_{\upsilon,0}$ and $C_{R,0}$ for $^{40}$K atoms as functions of the magnetic field magnitude changing from $191$G to $205$G across the laser-dressed $p$-wave resonance at a given temperature $T=6T_F$, spin polarization $P=0.1$, and $m_I=0$.
The red (dark gray) lines are calculated under the laser dressing, while the green (light gray) lines are calculated in the absence of the laser. The solid lines denote the results with coupling of $s$- and $p$-wave, the dashed lines denote the results of pure $p$-wave, and the dot-dashed black line denotes the result of pure $s$-wave. Parameters used for the plots are given in the main text.} \label{fig:ContactB}
\end{figure*}

For the numerical calculations, we take the atom density $n=1.50\times10^{13}{\rm  cm}^{-3}$~\cite{exp2004s}.
For $^{40}$K atoms,
we have the experimental parameters $\delta\mu_{m_I}=0.134\mu_{B}$ with the Bohr magneton $\mu_{B}$~\cite{exp2002s,exp2004s,exp2003p,review2010}.

The $s$-wave scattering length $a_s$ is given by~\cite{exp2002s,exp2004s,exp2003p,review2010}
\begin{eqnarray}
a_{s}=a_{bg}\left(1-\frac{\Delta_{s}}{B-B_{0,s}}\right),
\end{eqnarray}
where $B_{0,s}=202.1$G, $a_{bg}\simeq174a_{0}$, $a_{0}$ is the Bohr radius, and $\Delta_{s}\simeq8.0$G.

In absence of the laser field, the $p$-wave scattering volume can be conveniently calculated using~\cite{review2010,exp2004p,Yu2015exp}
\begin{eqnarray}
\tilde{\upsilon}_{m_I}
= \tilde{\upsilon}_{m_I}^{(bg)} \left( 1 - \frac{\Delta_{m_I} }{ B - B_{0,m_I} } \right),
\end{eqnarray}
where $\tilde{\upsilon}_{m_I=0}^{(bg)} = (101.6a_{0})^3$, $\tilde{\upsilon}_{m_I=\pm1}^{(bg)} = (96.74a_{0})^3$,
$\Delta_{m_I=0} = 21.95$G, $\Delta_{m_I=\pm1} = 24.99$G, $B_{0,m_I=0} = 198.8$G, and $B_{0,m_I=\pm1} = 198.3$G.

The $p$-wave effective range in absence of the laser field is~\cite{review2010,exp2004p,Yu2015exp}
\begin{eqnarray}
\frac{1}{\tilde{R}_{m_I}} = \frac{1}{\tilde{R}_{m_I}^{(bg)}} \left( 1 + \frac{ B - B_{0,m_I} }{ \Delta_{R,m_I} } \right),
\end{eqnarray}
where $\tilde{R}_{m_I=0}^{(bg)} = 47.19a_{0}$, $\tilde{R}_{m_I=\pm1}^{(bg)} = 46.22a_{0}$, $\Delta_{R,m_I=0} = -18.71$G, and $\Delta_{R,m_I=\pm1} = -22.46$G.

In presence of the laser field, we consider the typical experimental values $\gamma_{{\rm e}}=2\pi\times6$MHz, $\delta_{0} = -1.55$GHz, $\delta_{\pm1} = -2.90$GHz, $\Omega_{0}=2\pi\times57.14$MHz, $\Omega_{\pm1}=2\pi\times32.95$MHz, $B_{m_I=0} = 201.6$G, and $B_{m_I=\pm1} = 198.8$G~\cite{Zhangexp2017}.

Here, $B_{m_I=0} = 201.6$G is much closer to the $s$-wave Feshbach resonance $B_{0,s}=202.1$G than $B_{m_I=\pm1} = 198.8$G. Accordingly, we calculate the contacts $C_{\upsilon,m_I}$ and $C_{R,m_I}$ with $m_I=0$, for instance.

\subsection{Contacts}\label{7.1}

Figures~\ref{fig:ContactB}(a)-\ref{fig:ContactB}(c) show the contacts of $^{40}$K atoms as functions of the magnetic field magnitude changing from $191$G to $205$G across the laser-dressed $p$-wave resonance at a given temperature $T=6T_F\simeq2.1\mu$K~\cite{Onofrio1,Onofrio2}, spin polarization $P=0.1$, and $m_I=0$. The red (dark gray) lines are calculated under the laser dressing, while the green (light gray) lines are calculated in the absence of the laser. The solid lines denote the results with coupling of $s$- and $p$-wave, the dashed lines denote the results of pure $p$-wave, and the dot-dashed black line denotes the results of pure $s$-wave.

According to the laser-dressed $p$-wave interaction, the $s$-wave contact $C_{a}$ with laser dressing significantly decreases around the $p$-wave Feshbach resonance $198$G as shown in Fig.~\ref{fig:ContactB} (a). Such a behavior is a direct manifestation of few-body effects on the many-body level, and is useful for detecting the impact of dressing lasers on the system.

Second, Figs.~\ref{fig:ContactB}(b) and \ref{fig:ContactB}(c) show that the magnetic field points for the maximum values of $C_{\upsilon,0}$ and $|C_{R,0}|$ with laser are closer to the laser-dressed $p$-wave resonance $201.6$G than the corresponding results without laser, and the maximum values of $C_{\upsilon,0}$ and $|C_{R,0}|$ with laser are much larger than the corresponding results without laser.
This is according to the strong interplay of laser dressing and $p$-wave interaction.

Third, it is indicated from Figs.~\ref{fig:ContactB}(b) and \ref{fig:ContactB}(c) that the $p$-wave contacts $C_{\upsilon,0}$ and $|C_{R,0}|$ decrease more rapidly in the BEC limit under the influence of $s$-wave interaction, which is due to the interplay of $s$- and $p$-wave interactions on the many-body level.

\subsection{Spectral function}\label{7.2}

\begin{figure}
\includegraphics[width=8cm]{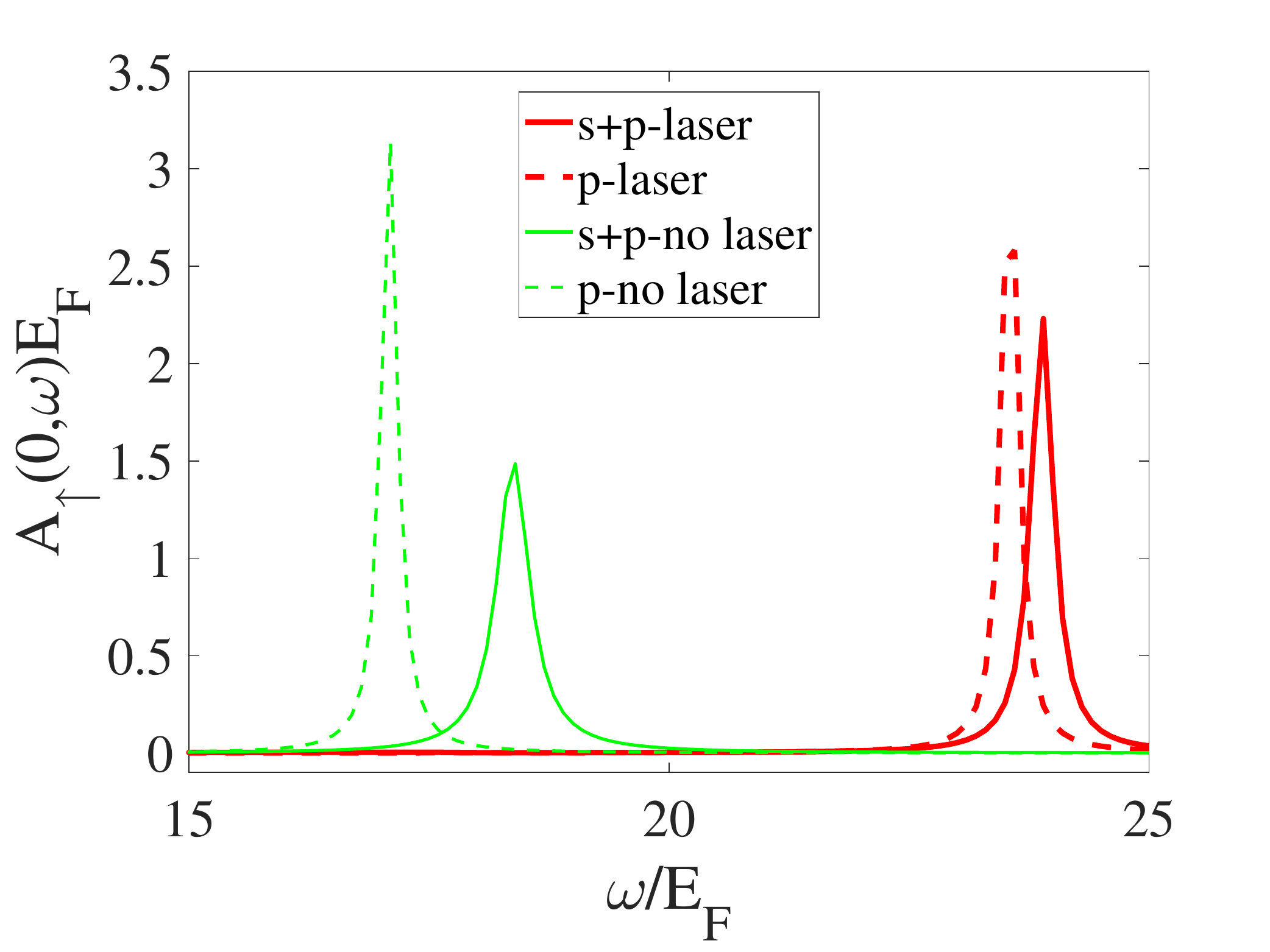}
\caption{(Color online) Spectral function $A_{\uparrow}(0,\omega)$ versus $\omega$ for $^{40}$K atoms at a given temperature $T=6T_F$, magnetic field magnitude $B=201$G, spin polarization $P=0.1$, and $m_I=0$.
Here, we choose $\theta=0$, where $\theta$ is the angle between momenta ${\bf p}$ and ${\bf p}'$ before and after the scattering event.
The line styles are similar to those in Fig.~\ref{fig:ContactB}. Parameters used for the plots are given in the main text.} \label{fig:spectral function}
\end{figure}

Figure~\ref{fig:spectral function} shows the spectral function of $^{40}$K atoms versus the frequency at a given temperature $T=6T_F$, magnetic field magnitude $B=201$G, spin polarization $P=0.1$, and $m_I=0$.

Similar to the contacts, the spectral function shows a very obvious laser-dressing effect on the many-body level.

\section{Summary}\label{8}

We have shown that, in a three-dimensional Fermi gas with laser-dressed mixed $s$- and $p$-wave interactions, the high-momentum tail of the density distribution can be characterized by a series of contacts which depend on the laser dressing. In particular, we find that the contact related to the velocity of the closed-channel molecules is anisotropic and the high-momentum tail of the momentum distribution show anisotropic behaviors of center-of-mass momentum. We then derive the universal relations, and numerically estimate the high-temperature contacts and spectral function which show the interplay of laser dressing and different partial-wave interactions on the many-body level.
In particular, the laser-dressing effect on the contacts and spectral function is visualized. The results here can be verified in current cold atom experiments.

\section*{Acknowledgements}

We thank Wei Yi, Liang-Hui Huang, Xiaoling Cui, Zhenhua Yu, Ming Gong, Pengfei Zhang, Mingyuan Sun, Lihong Zhou, Tian-Shu Deng, Lijun Yang, and Jing-Bo Wang for helpful discussions.
This work is supported by the National Key R\&D Program of China (Grant No. 2017YFA0304800) and the National Natural Science Foundation of China (Grant No. 11404106).
F.Q. acknowledges support from the project funded by the China Postdoctoral Science Foundation (Grant No. 2016M602011).

\appendix

\section{Functions in self-energy}\label{Appendix1}

The functions in the self-energy of Eqs.~(\ref{eq:self-energy1}) and (\ref{eq:self-energy2}) are given by
\begin{align}
F^{(s)}({\bf k},E_{\sigma})&= \int\frac{d^3{\bf P}}{(2\pi)^3} e^{-\beta\varepsilon_{{\bf P}-{\bf k}}}f^{(s)}\left(E_{\sigma} + \varepsilon_{{\bf P}-{\bf k}} - \frac{P^{2}}{4m}\right),
\end{align}
\begin{align}
H^{(s)}({\bf k},E_{\sigma})&= - \int\frac{d^3{\bf P}}{(2\pi)^3} e^{-\beta\frac{P^2}{4m}} \nonumber\\
&~~\times h^{(s)}\left(E_{\sigma} + \varepsilon_{{\bf P}-{\bf k}} - \frac{P^{2}}{4m}\right),
\end{align}
\begin{align}
f^{(s)}(z)&=\frac{\Theta(a_{s}^{-1})8\pi}{m^{2}a_{s}(z-E_{b,s})} + \frac{4}{m^{3/2}}\int_{0}^{\infty}\frac{\sqrt{x}dx}{(x-E_{b,s})(z-x)},
\end{align}
\begin{align}
h^{(s)}(z)&=\frac{\Theta(a_{s}^{-1})8\pi e^{-\beta E_{b,s}}}{m^{2}a_{s}(z-E_{b,s})} \nonumber\\
&~~+ \frac{4}{m^{3/2}}\int_{0}^{\infty}\frac{e^{-\beta x}\sqrt{x}dx}{(x-E_{b,s})(z-x)},
\end{align}
\begin{align}
&F^{(p)}({\bf k},E_{\sigma})= \sum_{m_I}16\pi Y_{1,m_I}({\bf \hat{p}})Y^{*}_{1,m_I}({\bf \hat{p}}')\nonumber\\
&~~\times \int\frac{d^3{\bf P}}{(2\pi)^3} e^{-\beta\varepsilon_{{\bf P}-{\bf k}}} f^{(m_I)}\left(E_{\sigma} + \varepsilon_{{\bf P}-{\bf k}} - \frac{P^{2}}{4m}\right),
\end{align}
\begin{align}
&H^{(p)}({\bf k},E_{\sigma})= - \sum_{m_I}16\pi Y_{1,m_I}({\bf \hat{p}})Y^{*}_{1,m_I}({\bf \hat{p}}') \nonumber\\
&~~\times \int\frac{d^3{\bf P}}{(2\pi)^3} e^{-\beta\frac{P^2}{4m}} h^{(m_I)}\left(E_{\sigma} + \varepsilon_{{\bf P}-{\bf k}} - \frac{P^{2}}{4m}\right),
\end{align}
\begin{align}
&f^{(m_I)}(z)=\frac{-\Theta(\upsilon_{m_I}^{-1})\pi}{(mR_{m_I}^{-1})^{2}\upsilon_{m_I}(z-E_{b,m_I})}\nonumber\\
&~~ + \int_{0}^{\infty}\frac{m^{3/2}x^{5/2}dx}{[(\upsilon_{m_I}^{-1}+R_{m_I}^{-1}mx)^2+(mx)^3](z-x)},
\end{align}
\begin{align}
&h^{(m_I)}(z)=\frac{-\Theta(\upsilon_{m_I}^{-1})\pi e^{-\beta E_{b,m_I}}}{(mR_{m_I}^{-1})^{2}\upsilon_{m_I}(z-E_{b,m_I})}\nonumber\\
&~~ + \int_{0}^{\infty}\frac{e^{-\beta x}m^{3/2}x^{5/2}dx}{[(\upsilon_{m_I}^{-1}+R_{m_I}^{-1}mx)^2+(mx)^3](z-x)}.
\end{align}

Here $E_{b,s}=-1/(ma^2_s)$ and $E_{b,m_I}=-R_{m_I}/(m\upsilon_{m_I})$ are the $s$- and $p$-wave two-body bounding energies, respectively.

\end{CJK*}
\end{document}